\documentclass[12pt,letterpaper,english]{article}

 \usepackage{setspace,graphics}
 \usepackage[dvips]{epsfig} 
 \usepackage{a4,amssymb,epsfig,array,cite}
 \usepackage[hypertex]{hyperref}

\usepackage{amsfonts,bm,amssymb,euscript,array,babel}

\newcounter{multieqs}

\newcommand{\be}{\begin{equation}}
\newcommand{\ee}{\end{equation}}
\newcommand{\eq}[1]{(\ref{#1})}

\def\nn{\nonumber}
\def\bea{\begin{eqnarray}}
\def\eea{\end{eqnarray}}
\def\obar{\overline}

\def\beqa{\begin{eqnarray}} 
\def\eeqa{\end{eqnarray}} 
\def\beq{\begin{equation}} 
\def\eeq{\end{equation}}

\def\Tr{{\rm Tr}}
\def\rot{{\vec\nabla\times}}
\def\div{{\vec\nabla\cdot}}

\def\a{\alpha}          
\def\b{\beta}           
\def\c{\gamma}    
\def\d{\delta}

\def\g{\gamma}

 \def\L{\Lambda} \def\la{\lambda}
\def\m{\mu}

\def\cA{{\cal A}}  \def\cC{{\cal C}}
  
 \def\cH{{\cal H}} 
  
\def\cM{{\cal M}} \def\cN{{\cal N}}

\def\R{{\mathbb R}}

\def\one{\mbox{1 \kern-.59em {\rm l}}}

\def\bit{\begin{itemize}}
\def\eit{\end{itemize}}

\def\({\left(}
\def\){\right)}
\def\diag{\mbox{diag}}

\def\d{\delta}

 \def\del{\partial}

\def\uno{\mbox{1 \kern-.59em {\rm l}}}

\def\bcomment#1{}

\renewcommand{\title}[1]{\vspace{10mm}\noindent{\Large{\bf #1}}\vspace{8mm}}
\newcommand{\authors}[1]{\noindent{\large #1}\vspace{5mm}}
\newcommand{\address}[1]{{\itshape #1\vspace{2mm}}}

\begin{document}

\begin{titlepage}

\begin{flushright}
UWTHPh-2008-11\\
arXiv:0806.2032[hep-th]
\end{flushright}

\begin{center}
  
\title{Emergent Gravity and Noncommutative Branes \\[1ex]
from Yang-Mills Matrix Models}

\authors{Harold {\sc Steinacker}${}^{1}$}

\address{ Fakult\"at f\"ur Physik, Universit\"at Wien\\
Boltzmanngasse 5, A-1090 Wien, Austria}

\footnotetext[1]{harold.steinacker@univie.ac.at, phone: +43 1 4277
  51526, fax +43 1 4277 9515}

\vskip 2cm

\textbf{Abstract}

\vskip 3mm 

\begin{minipage}{14cm}%

The framework of emergent gravity arising 
from Yang-Mills matrix models is developed further,
for general noncommutative branes embedded in  $\R^D$. 
The effective metric on the brane
turns out to have a universal form reminiscent of the open string
metric, depending on the dynamical Poisson structure 
and the embedding metric in  $\R^D$. 
A covariant form of the tree-level equations of motion 
is derived, and the Newtonian limit
is discussed. This points to the necessity of branes
in higher dimensions. The quantization 
is discussed qualitatively, which singles out the IKKT model as a prime
candidate for a quantum theory of gravity coupled to matter. 
The Planck scale is then identified with the scale of 
$N=4$ SUSY breaking.  
A mechanism for avoiding the cosmological constant problem 
is exhibited.

\end{minipage}

\end{center} 
\vspace{0.5cm}



\end{titlepage}

\setcounter{page}0
\thispagestyle{empty}

\begin{spacing}{.3}
{
\noindent\rule\textwidth{.1pt}            
   \tableofcontents
\vspace{.6cm}
\noindent\rule\textwidth{.1pt}
}
\end{spacing}


\section{Introduction}

The notion of space-time which underlies the presently accepted models
of fundamental matter and interactions goes back to Einstein.
Space-time is modeled by a 4-dimensional
manifold, whose geometry is determined by a metric with
Lorentzian signature. 
This notion escaped the quantum revolution essentially
unchanged, even though Quantum Mechanics combined with
General Relativity strongly suggests a ``foam-like'' or quantum
structure at the Planck scale.
While some kind of quantum structure of space-time indeed arises
e.g. in string theory or loop quantum gravity, a satisfactory
understanding is still missing.

A different approach to this problem 
has been pursued in recent years, starting
with some explicit quantization of space-time and
attempting to construct physical models on such a background. 
The classical space-time $\R^4$ is replaced by a quantized 
or ``noncommutative'' (NC) space, 
where the coordinate functions $x^\mu$ satisfy 
nontrivial commutation relations such as
$[x^\mu,x^\nu] = i \theta^{\mu\nu}$.
This leads to non-commutative field theory, see e.g.
\cite{Doplicher:1994tu,Douglas:2001ba,Szabo:2001kg}. 
At the semi-classical level, 
these commutation relations determine 
a Poisson structure $\theta^{\mu\nu}$ on space-time,
which is fixed by construction. 
However, since quantized spaces are expected to arise 
from quantum gravity, 
it seems more appropriate to consider a {\em dynamical}
Poisson structure at the semi-classical level. A straightforward 
generalization of General Relativity is then
inappropriate; indeed any quantum structure of space-time 
rules out classical intuitive principles.
Rather, one should look for simple models of dynamical noncommutative
(or Poisson) spaces, with the hope that they will effectively
incorporate gravity.

Such models are indeed available and known as Matrix-models of 
Yang-Mills type. They have the form 
$S = Tr [X^a,X^b][X^{a'},X^{b'}] \d_{aa'} \d_{bb'} + ... $,
where indices run from 1 to $D$. It is well known that 
these models admit noncommutative spaces (``NC branes'') as solutions,
such as the Moyal-Weyl quantum plane $\R^4_\theta$; see e.g.
\cite{Aoki:1999vr,Kabat:1997im,Myers:1999ps,
Alekseev:2000fd,Nair:1998bp,Banks:1996nn,Azuma:2002zi}. 
However, most of the work up to now is focused on special
NC branes with a high degree of symmetry.
For generic NC spaces with non-constant $\theta^{\mu\nu}(x)$, 
it was shown in \cite{Steinacker:2007dq} that 
the kinetic term for any ``field'' coupled 
to the $D=4$ matrix model is governed by an effective metric 
$\tilde G^{ab}(x) = \rho\, \theta^{aa'}(x) \theta^{bb'}(x) \d_{a'b'}$,
including nonabelian gauge fields.
This nicely explains the observed relation in \cite{Rivelles:2002ez}
between NC $U(1)$ gauge fields and gravitational degrees of freedom,
see also \cite{Muthukumar:2004wj,Banerjee:2004rs,Fatollahi:2008dg} for related work.
Since this effective metric is dynamical, 
these YM Matrix Models contain effectively some version of gravity,
thus realizing the idea that gravity should emerge from NC gauge theory 
\cite{Yang:2006mn,Rivelles:2002ez}.
As argued in \cite{Steinacker:2007dq}, 
an effective action for gravity is induced upon quantization, 
with the remarkable
feature that the ``would-be cosmological term'' decouples 
from the model due to the constrained class of 
metrics. This makes the mechanism of induced gravity
feasible at the quantum level, 
and suggests that the Newton constant resp. the Planck scale
is related to an effective UV-cutoff of the model.  
A detailed analysis taking into account UV/IR mixing \cite{Grosse:2008xr}
and fermions \cite{Klammer:2008df} 
singles out the $N=4$ supersymmetric
extensions of the model, where such a cutoff is given by the
scale of  $N=4$ SUSY breaking. This amounts to  $D=10$,
which is nothing but the IKKT model
\cite{Ishibashi:1996xs},  originally proposed as 
a nonperturbative definition of IIB string theory\footnote{
As such, the presence of gravity in this model 
is expected and to some extent verified, 
cf.  \cite{Banks:1996vh,Kabat:1997sa,
Ishibashi:1996xs,Bigatti:1997jy,Aoki:1999vr,Ishibashi:2000hh,
Kimura:2000ur}. 
However, what is usually considered are 
effects of  D=10 (super)gravity, modeled by  
interactions of separated ``D-objects'', represented 
by block-matrices. In contrast, emergent NC gravity 
describes interactions within (generic) NC branes in this model.
Evidence for gravity on simple NC branes was obtained previously
in \cite{Ishibashi:2000hh,Kitazawa:2006pj}.}.

In the present paper, we develop the framework for 
emergent gravity on general NC branes with nontrivial embedding in $\R^D$.
This works out very naturally, leading to a
simple generalization of the effective metric which is
strongly reminiscent of the open string metric \cite{Seiberg:1999vs}, 
involving the general Poisson tensor  
and the embedding metric.
We establish in Section \ref{sec:basic} the relevant geometry, 
find the semi-classical form of the bare matrix-model action for 
general NC branes in 
$\R^D$, and obtain covariant equations of motion. 
This generalizes the well-known case of flat or 
highly symmetric branes to the generic case, and shows how the 
would-be $U(1)$ gauge field is absorbed in the effective metric on the
brane. In Section \ref{sec:linearized}, the Newtonian limit 
of emergent gravity
is studied in detail. It turns out that
even though it is possible to reproduce the Newtonian potential for
general mass distributions, the relativistic corrections 
are in general not correctly reproduced in $D=4$ matrix models.
This provides one motivation to consider general branes embedded in 
higher-dimensional matrix models, which admit a 
much richer class of geometries and promise to overcome this problem.
The compactification of higher-dimensional NC branes
is described in Section \ref{sec:compactification} 
with the example of fuzzy spheres in extra dimensions. 

Higher dimensions, more precisely $D=10$ resp. $N=4$ SUSY
also appears to be 
required by consistency at the quantum level. From the point of 
view of emergent gravity, this condition arises as a 
result of UV/IR mixing in NC gauge theory. 
This is discussed qualitatively
in Section \ref{sec:induced-grav} 
along the lines of \cite{Steinacker:2007dq}, 
leading to an induced gravity action.
In Section \ref{sec:departures} 
some differences to General Relativity are discussed,
most notably the presence of intrinsic scales and preferred
coordinates, as well as the different role of the 
``would-be cosmological constant term''.
As an illustration of the formalism, 
we also give a (unphysical) 
solution of the bare equations of motion in Section 
\ref{sec:spherical}. Finally, a 
matrix version of a conserved energy-momentum tensor
is derived.

The results of this paper provide a rich framework for
the search of realistic solutions of emergent NC gravity.
The main missing piece is the analog of the Schwarzschild solution,
which is nontrivial because
the quantum effective action at least at one loop must be 
taken into account. But in any case, it is clear that 
these models do contain a version of gravity
in an intrinsically noncommutative way, and they
have a good chance to be well-defined at the quantum level
at least for the IKKT model. This certainly provides motivation 
for a thorough investigation.

\section{The Matrix Model}
\label{sec:basic}

Consider the matrix model with action 
\be
S_{YM} = - Tr [X^\mu,X^\nu] [X^{\mu'},X^{\nu'}] g_{\mu\mu'}g_{\nu\nu'},
\label{YM-action-1}
\ee
for
\be
g_{\mu\mu'} = \delta_{\mu\m'} \quad \mbox{or}
 \quad g_{\mu\mu'} = \eta_{\mu\mu'} 
\label{background-metric}
\ee
in the Euclidean  resp.  Minkowski case.
The "covariant coordinates"  $X^\mu,  \,\, \mu=1,2,3,4$ 
are hermitian matrices 
or operators acting on some Hilbert space $\cH$.
We will denote the commutator of 2 matrices as 
\be
[X^\mu,X^\nu] = i \theta^{\mu\nu}
\label{theta-def}
\ee
so that $\theta^{\mu\nu} \in L(\cH)$ are antihermitian\footnote{in
contrast to the conventions in \cite{Steinacker:2007dq}} 
matrices, which are 
{\em not} assumed to be proportional to $\one_\cH$.
We focus here on configurations $X^\mu$ 
 which can be interpreted as quantizations 
of coordinate functions $x^\mu$ on
a Poisson manifold $(\cM,\theta^{\mu\nu}(x))$ with general
Poisson structure $\theta^{\mu\nu}(x)$. This defines the geometrical 
background under consideration, and conversely 
essentially any Poisson manifold provides (locally) a possible
background $X^\mu$ \cite{Kontsevich:1997vb}.
More formally, this means that there is an isomorphism of vector spaces
\be
\begin{array}{ccl}
\cC(\cM) &\to& \cA\,\subset \, L(\cH)\, \\
 f(x) &\mapsto& \hat f(X) \\
i\{f,g\} &\mapsto& [\hat f,\hat g] + O(\theta^2)
\end{array}
\label{map} 
\ee
Here $\cC(\cM)$ denotes some space of functions on $\cM$, 
and $\cA$ is interpreted as quantized algebra of 
functions\footnote{Roughly speaking $\cA$ is the algebra 
generated by $X^\mu$, but technically one usually considers some 
subalgebra corresponding to well-behaved functions.} 
 on $\cM$. This allows to replace
$[\hat f(X),\hat g(X)] \to i  \{f(x),g(x)\}$ to leading order in $\theta^{\mu\nu}$.
In particular, we can then write 
\be
[X^\mu,f(X)] \sim i\theta^{\mu\nu}(x) 
\frac{\partial}{\partial x^\nu} f(x)
\label{derivation}
\ee
which will be used throughout this paper, denoting with
$\sim$  the leading contribution in a semi-classical 
expansion in powers of $\theta^{\mu\nu}$.

In order to derive the effective metric on $\cM$,
let us now consider a scalar field 
coupled to the matrix model \eq{YM-action-1}. The only possibility 
to write down kinetic terms for matter fields is through commutators 
$[X^\mu,\Phi]$ using \eq{derivation}. Thus consider the action
$S = S_{YM} + S[\Phi]$ where
\bea
S[\Phi] &=& - Tr\, g_{\mu\mu'} 
[X^\mu,\Phi][X^{\mu'},\Phi]  \nn\\
&\sim& \frac 1{(2\pi)^2}\,\int d^4 x\, \rho(x)\,  
G^{\mu\nu}(x)\,\frac{\partial}{\partial x^\mu}\Phi(x) 
\frac{\partial}{\partial x^\nu} \Phi(x) .
\label{scalar-action-0}
\eea
Here 
\be
G^{\mu\nu}(x) = \theta^{\mu\mu'}(x) \theta^{\nu\nu'}(x)\, 
g_{\mu'\nu'} \, 
\label{effective-metric}
\ee
is interpreted as metric on $\cM$ in $x$ coordinates.
We will assume in this paper that $\theta^{\mu\nu}(x)$ is nondegenerate.
Then the symplectic measure on $(\cM,\theta^{\mu\nu}(x))$
is given by the scalar density
\be
\rho(x) \equiv |\theta^{-1}_{\mu\nu}(x)|^{1/2}
= |G_{\mu\nu}(x)|^{1/4} |g_{\mu\nu}|^{1/4} 
\equiv \L_{NC}^4(x)\, ,
\label{rho-def-flat}
\ee
which can be interpreted as ``local'' non-commutative scale $\L_{NC}$.
In the preferred $x$ coordinates characterized by
\eq{background-metric}, $\rho(x)$ coincides with the 
dimensionless scalar function
\be
e^{-\sigma} 
= \frac{|G_{\mu\nu}(x)|^{1/4}}{|g_{\mu\nu}(x)|^{1/4}} 
= \frac{|\theta^{-1}_{\mu\nu}(x)|^{1/2}}{|g_{\mu\nu}|^{1/2}} 
\, .
\label{rho-tilde-def-flat}
\ee
The action \eq{scalar-action-0} can now be written 
in a covariant manner as
\be
S[\Phi] = \frac 1{(2\pi)^2}\,\int d^4 x\, 
\tilde G^{\mu\nu}(x)\,
 \partial_{\mu}\Phi(x) \partial_{\nu}\Phi(x) 
= \frac 1{(2\pi)^2}\,\int d^4 x\, \sqrt{|\tilde G_{\mu\nu}|}\,\, 
\Phi(x)\Delta_{\tilde G} \Phi(x) \, ,
\label{scalar-action-geom}
\ee
Here $\Delta_{\tilde G}$
is the Laplacian for the  metric \cite{Steinacker:2007dq}
\bea
\tilde G^{\mu\nu}(x) &=& |G_{\mu\nu}|^{1/4}\, G^{\mu\nu}(x) 
= e^{-\sigma}\,G^{\mu\nu}(x) ,\nn\\
|\tilde G^{\mu\nu}| &=& 1\, ,
\label{metric-unimod}
\eea
which is unimodular in the preferred $x^\mu$ coordinates.
By definition, 
$\tilde G^{\mu\nu}(x)$ is the effective metric for the scalar field.
Because it
enters in the kinetic term for any matter coupled to the
matrix model,
it plays the role of a gravitational metric
\be
d s^2 = \tilde G_{\mu\nu}(x)\, dx^\mu dx^\nu \, .
\label{ds-2}
\ee
Up to  certain density factors,
this also applies to nonabelian gauge fields 
as shown in \cite{Steinacker:2007dq} and for fermions 
\cite{Klammer:2008df}. 
Therefore the Poisson manifold under consideration naturally 
acquires a metric structure 
$(\cM,\theta^{\mu\nu}(x),\tilde G^{\mu\nu}(x))$, 
which is determined by the Poisson structure and the 
flat background metric $g_{\mu\nu}$.

\paragraph{Equations of motion.}

The basic matrix model action \eq{YM-action-1} leads to the
e.o.m. for $X^\mu$ 
\bea
[X^\mu,[X^{\mu'},X^{\nu'}]] g_{\mu\mu'} = 0 \, .
\label{eom-Y}
\eea
This can be written in the semi-classical limit as
$\theta^{\mu \g} \partial_\g \theta^{\mu'\nu} g_{\mu\mu'} = 0$, or
\be
G^{\g\eta}(x)\, \partial_\g \theta^{-1}_{\eta \nu}= 0 \, .
\label{eom-Y-2}
\ee
These equations are not covariant, they are valid only in the
coordinates $x^\mu$ where the ``background metric'' $g_{\mu\nu}$ 
in the matrix model is either $\delta_{\mu\nu}$ or $\eta_{\mu\nu}$.
As shown in Appendix B, these equations of motion can be written 
in a covariant manner as 
\be
\fbox{$
\tilde G^{\eta\g}(x)\, \tilde\nabla_{\g} 
(e^{\sigma} \theta^{-1}_{\eta \nu}) 
\, = e^{-\sigma} \, \tilde G_{\mu\nu}\theta^{\mu \g}\,\partial_\g\eta(x)
$}
\label{eom-geom-covar-1}
\ee
where
\bea
\eta(x) &=& \frac 14\, G^{\mu\nu} g_{\mu\nu} 
= \frac 14\, G^{\mu\nu} G^{\mu'\nu'}
\theta^{-1}_{\mu\mu'} \theta^{-1}_{\nu\nu'}  \label{eta-def-1}
\eea
and $\tilde \nabla$ denotes the Levi-Civita connection with respect to
the metric $\tilde G^{\mu\nu}$.
Note that the ``background'' metric $g_{\mu\nu}$ 
is absorbed completely. \eq{eom-geom-covar-1} can be written as 
\bea
\tilde G^{\g \eta}(x)\, \tilde\nabla_\g  \theta^{-1}_{\eta \nu} 
 &=&  - \tilde G^{\g\eta}(x)\,\theta^{-1}_{\eta\nu} \partial_\g \sigma
+ e^{-2\sigma}\,\tilde G_{\mu\nu}\theta^{\mu \g}\,\partial_\g\eta(x)\, ,
\label{eom-tensor}
\eea
which has the form of
covariant Maxwell equations with source. 
The obvious advantage of this covariant form of the equations of
motion is that we can now use any
adapted coordinates, in particular rotation-symmetric ones etc. This
should help to find solutions. Nevertheless, this 
should not obscure the fact that the underlying matrix model is {\em not}
invariant under diffeomorphisms: the background
metric $g_{\mu\nu}$ is constant, and there is no obvious way to
transform it at the level of the matrix model. Only in the
semi-classical limit we can allow general coordinates and rewrite
things in a coordinate independent way, at the expense of
introducing
a flat background metric $g_{\mu\nu}$. 

In principle of course, the equation of motion for $X^\mu$
is modified due to the presence of the scalar field.
However for small coupling or energy, we can presumably neglect this
back-reaction of matter on the geometry. 
It will be taken into account in section 
\eq{sec:extradim}.

The equation of motion for the scalar field $\phi$ are
\be
0 \, = \, [X^\mu,[X^\nu,\phi]] g_{\mu\nu} \, \sim \, 
\theta^{\mu\mu'} \partial_{\mu'} (\theta^{\nu\nu'} 
\partial_{\nu'} \phi)  g_{\mu\nu} \,.
\ee
As shown in Appendix A, 
this can be written as 
\be
\Delta_{\tilde G} \phi = 
(\tilde G^{\mu\nu}\partial_{\mu}  \partial_{\nu} -
\tilde\Gamma^\mu \partial_\mu) \phi =0
\label{eom-phi-4D}
\ee
where $\tilde \Gamma^\mu = \tilde G^{\nu\eta}\, 
\tilde\Gamma_{\nu\eta}^\mu$
and $\tilde\Gamma_{\nu\eta}^\mu$ are the Christoffel symbols of
$\tilde G_{\mu\nu}$. This follows also immediately from 
the covariant
form \eq{scalar-action-geom} of the scalar action.
We will show moreover in Appendix A that in the preferred
$x^\mu$ coordinates defined by the matrix model, 
the equation of 
motion \eq{eom-Y-2} for $X^\mu$ resp. $\theta^{-1}_{\mu\nu}$ 
is equivalent to the non-covariant equation
\eq{tilde-Gamma-vanish}
\be
\tilde \Gamma^\mu =0 \, .
\ee
In these coordinates, the equation of motion for $\phi$
takes the simple form 
$\tilde G^{\mu\nu}\partial_{\mu} \partial_{\nu}
\phi =0$, for on-shell geometries.

The semi-classical form of the matrix model action \eq{YM-action-1}
is 
\be
S_{YM} = \frac 4{(2\pi)^2}\,\int d^4x\,\rho(x) \eta(x) \,.
\label{S-simple}
\ee
The equation of motion 
\eq{eom-geom-covar-1} can be derived directly
from this action.
We will give this derivation in the next section,
in the context of general branes embedded in $\R^D$.

\subsection{Noncommutative branes and extra dimensions}
\label{sec:extradim}

Let us discuss scalar matter from the point of view of extra dimensions.
Recall that e.g. the action for a scalar field is given by 
additional terms of the type
\be
Tr [X^\mu,\phi] [X^\nu,\phi] \eta_{\mu\nu}\, .
\ee
The combined action can be interpreted as matrix model 
with extra dimensions, where
one coordinate denoted as $\phi$ is a function of the other 4
coordinates. Therefore we consider more generally 
\be
S_{YM} = - Tr [X^a,X^b] [X^{a'},X^{b'}] 
\eta_{aa'}\eta_{bb'},
\label{YM-action-extra}
\ee
for hermitian matrices 
or operators  $X^a,  \,\, a=1,..., D$ 
acting on some Hilbert space $\cH$.
To avoid a proliferation of symbols we fix the background 
to have the Minkowski metric; 
the Euclidean case is completely parallel, 
replacing $\eta_{ab}$ with  $\d_{ab}$.
A scalar field can therefore be interpreted
as defining an embedding of a 4-dimensional manifold 
(a ``3-brane'')
in a higher-dimensional space.  
This naturally suggests to consider a higher-dimensional version of the
Yang-Mills matrix model, such as the IKKT model in 10 dimensions.

We want to consider general $2n$ -dimensional 
noncommutative spaces $\cM_\theta^{2n} \subset \R^{D}$
(a $2n-1$ brane) in $D$ dimensions. 
We correspondingly split the matrices as
\be
X^a = (X^\mu,\phi^i), \qquad \mu = 1,...,2n, 
\,\,\, i=1, ..., D-2n.
\label{extradim-splitting}
\ee
The basic example is a flat embedding of a
4-dimensional NC background with
\bea
[X^\mu,X^\nu] &=& i\theta^{\mu\nu}, \qquad \mu, \nu= 1,...,4, \nn\\
\phi^i &=& 0,\qquad\quad\,  i=1, ..., D-4
\eea
where $X^\mu$ generates a 4-dimensional NC brane $\cM^4_\theta$. 
Then the discussion of the previous section applies, and 
fluctuations of $\phi^i(x)$ can be interpreted as scalar fields on
$\cM^4_\theta$.
More generally, we can interpret $\phi^i(x)$ as defining 
the embedding of a $2n$ - dimensional submanifold $\cM^{2n} \subset
\R^{D}$, equipped with a nontrivial induced metric. The support
(``D-dimensional spectrum'') of $X^a \sim x^a$ 
will then be concentrated on $\cM^{2n} \subset \R^{D}$ in the
semi-classical limit.
Expressing the $\phi^i$ in terms of $X^\mu$, we obtain
\be
[\phi^i,f(X^\mu)] \,\sim\, i\theta^{\mu\nu}\partial_\mu \phi^i \partial_\nu f
= ie^{\mu}(f)\, \partial_\mu \phi^i
\ee
in the semi-classical limit.
This involves only the components $\mu = 1, ..., 2n$ of the
antisymmetric tensor $[X^a,X^b] \sim i\theta^{ab}(x)$, which has rank $2n$
in this case.
Here
\be
e^{\mu} := -i[X^\mu,.] \,\sim\, \theta^{\mu\nu} \partial_\nu 
\ee
are derivations, which span the tangent space of $\cM^{2n}\subset \R^{D}$.
They will define a preferred frame below. We can then interpret 
\be
[X^\mu,X^\nu] \sim i\theta^{\mu\nu}(x)
\label{theta-induced}
\ee
as Poisson structure on $\cM^{2n}$ (assuming that it is
no-degenerate), noting that the Jacobi identity
is trivially satisfied. This is the Poisson structure 
 on $\cM^{2n}$ whose quantization is given by the matrices 
$X^\mu, \,\, \mu = 1,..., 2n$, interpreted as quantization of 
the  coordinate functions $x^\mu$ on $\cM^{2n}$. 
Conversely, any $\theta^{\mu\nu}(x)$ \eq{theta-induced}
can be (locally) quantized, and provides together with arbitrary
$\phi^i(x)$ a quantization of $\cM^{2n} \subset \R^D$ 
as described above.
Note that this Poisson structure is defined intrinsically by the 
configurations of the matrix model,
independent of the choice\footnote{For
generic embeddings, the separation \eq{extradim-splitting}
is arbitrary, and we are free to choose  different $2n$ components
among the $\{X^a\}$ as generators of tangential vector fields.
This is a particular change of coordinates on $\cM^{2n}$, 
which from the field theory point of view corresponds
to a remarkable transformation exchanging
fields with coordinates, reminiscent of T-duality in string theory.
In any case, note that $i\theta^{\mu\nu}$ is not naturally a pull-back
of some non-degenerate Poisson or symplectic structure on $\R^D$.}
 made in \eq{extradim-splitting}.
Assuming that $\theta^{\mu\nu}(x)$ is non-degenerate,
we denote its inverse matrix with
\be
\theta^{-1}_{\mu\nu}(x)\, ,
\ee
which defines a symplectic form on $\cM^{2n}$. 
Finally, 
the trace is again given semi-classically by the 
volume of the symplectic form,
\be
(2\pi)^{n}\, Tr f \sim \int d^{2n} x\, \rho(x)\, f
\ee
with $\rho = (\det \theta^{-1}_{\mu\nu})^{1/2}$ 
generalizing \eq{rho-def-flat}.

We are now in a position to extract the semi-classical limit
of the matrix model and its physical interpretation.
To understand the effective geometry on $\cM^{2n}$,
consider again a (test-) particle on $\cM^{2n}$, 
modeled by some additional
scalar field $\varphi$
(this could be e.g. $su(k)$ components of $\phi^i$).
The kinetic term due to the matrix model must have the form
\bea
S[\varphi] &\equiv& - Tr [X^a,\varphi][X^b,\varphi] \eta_{ab} = 
- Tr \([X^\mu,\varphi][X^\nu,\varphi] \eta_{\mu\nu} 
  + [\phi^i,\varphi][\phi^j,\varphi] \delta_{ij}\) \nn\\
&\sim&  Tr \,\(\theta^{\mu\mu'} \theta^{\nu\nu'}\, \partial_{\mu'} \varphi
   \partial_{\nu'} \varphi \eta_{\mu\nu} 
 + \theta^{\mu\mu'} \theta^{\nu\nu'}\, 
\partial_{\mu}\phi^i \partial_{\mu'}\varphi
  \, \partial_{\nu}\phi^j \partial_{\nu'}\varphi \,\delta_{ij}\) \nn\\
&=&  Tr \, \theta^{\mu\mu'} \theta^{\nu\nu'}\, 
 \(\eta_{\mu\nu} +  \partial_{\mu}\phi^i \partial_{\nu}\phi^j \delta_{ij} \) 
 \partial_{\mu'} \varphi \partial_{\nu'} \varphi \nn\\
&\sim&  \frac 1{(2\pi)^n}\,\int d^{2n} x\; \rho(x) \, G^{\mu\nu}(x)
 \partial_{\mu} \varphi \partial_{\nu} \varphi  
\label{covariant-action-scalar-0}
\eea
where
\bea
g_{\mu\nu}(x) &=& \eta_{\mu\nu} 
+  \partial_{\mu}\phi^i \partial_{\nu}\phi^j\delta_{ij} 
\, = \, \partial_\mu x^a\partial_\nu x^b\, \eta_{ab} \label{g-def}\\
G^{\mu\nu}(x) &=& \theta^{\mu\mu'}(x) \theta^{\nu\nu'}(x) 
 g_{\mu'\nu'}(x) \label{G-def-general}\\
\rho(x) &=& |\theta^{-1}_{\mu\nu}|^{1/2} = |G_{\mu\nu}|^{1/4} |g_{\mu\nu}(x)|^{1/4} .
\label{rho-def-general}
\eea
Here $g_{\mu\nu}(x)$ 
is the metric induced on $\cM^{2n}\subset \R^{D}$ via 
pull-back 
of $\eta_{ab}$ on $\R^{D}$. Now $g_{\mu\nu}(x)$ 
is no longer flat in
general. So far, the kinetic term
does not quite have the correct covariant form. 
This can be achieved by a 
suitable rescaling of $G^{\mu\nu}(x)$: 
generalizing the corresponding quantities in 
\eq{rho-tilde-def-flat} and \eq{metric-unimod}, we define 
\bea
\tilde G^{\mu\nu}(x) &=& e^{-\sigma}\, G^{\mu\nu}(x) \label{G-def-general-tilde}\nn\\
\rho\, G^{\mu\nu} &=& |\tilde G_{\mu\nu}|^{1/2}\,
\tilde G^{\mu\nu}(x) \label{covariant-rescaling}\\
e^{-(n-1)\sigma} &=& |G_{\mu\nu}|^{1/4}|g_{\mu\nu}(x)|^{-\frac 14}\, 
   = \rho\, |g_{\mu\nu}(x)|^{-\frac 12} \nn\\
|\tilde G_{\mu\nu}| 
&=& |\theta^{-1}_{\mu\nu}|^{\frac{n-2}{n-1}}\,
|g_{\mu\nu}(x)|^{\frac{1}{n-1}}
\label{G-tilde-general}
\eea
Then the action \eq{covariant-action-scalar-0} 
has the correct covariant form
\be
S[\varphi] =\frac 1{(2\pi)^n}\, \int d^{2n} x\; 
|\tilde G_{\mu\nu}|^{1/2}\,\tilde G^{\mu\nu}(x)
 \partial_{\mu} \varphi \partial_{\nu} \varphi  \,.
\label{covariant-action-scalar}
\ee
Therefore the kinetic term on  $\cM^{2n}_\theta$
is governed by the metric $\tilde G^{\mu\nu}(x)$, 
which has almost the same form as \eq{metric-unimod}
except that the constant background metric $g_{\mu\nu}$ is now replaced
by the induced metric $g_{\mu\nu}(x)$ on $\cM^{2n}\subset \R^{D}$.
The matrix model action \eq{YM-action-extra} 
can be written in the semi-classical limit as 
\be
S_{YM} = - Tr[X^a,X^b][X^{a'},X^{b'}] \eta_{aa'} \eta_{bb'} 
\,\sim\, \frac 4{(2\pi)^n}\,\int d^{2n} x\,  \rho(x) \eta(x) ,
\label{S-semiclassical-general}
\ee
where
\bea
4 \eta(y) &=&  G^{\mu\nu}(x) g_{\mu\nu}(x) 
=  (\eta_{\mu\nu} 
+ \partial_\mu\phi_r^i\partial_\nu\phi_r^j \d_{ij})
\theta^{\mu\mu'}\theta^{\nu\nu'}
(\eta_{\mu'\nu'} 
+ \partial_{\mu'}\phi_s^{i'}\partial_{\nu'}\phi_s^{j'} \d_{i'j'}) \nn\\
&=&  \(\theta^{\mu\mu'}\theta^{\nu\nu'}\eta_{\mu'\nu'}
\eta_{\mu\nu} + 2 \theta^{\mu\mu'}\theta^{\nu\nu'} \eta_{\mu'\nu'}
\partial_\mu\phi^i \partial_\nu \phi^{i'} \d_{ii'} 
 + \theta^{\mu\eta}\partial_\mu\phi^i\partial_\eta\phi^j 
\theta^{\mu'\eta'}\partial_{\mu'}\phi^{i'}\partial_{\eta'}\phi^{j'}
\d_{ii'}\d_{jj'}\) \nn\\
&\sim& - [X^a,X^b][X^{a'},X^{b'}] \eta_{aa'} \eta_{bb'}
\label{eta-def}
\eea
generalizes \eq{eta-def-1}.

There are 2 interesting 
special cases. For 4-dimensional
NC spaces, we have 
\be
|\tilde G_{\mu\nu}(x)| = |g_{\mu\nu}(x)|, \qquad 2n=4
\label{G-g-4D}
\ee
which means that the Poisson tensor $\theta^{\mu\nu}$ does 
not enter the Riemannian volume at all. This provides a
very interesting mechanism for ``stabilizing flat space'', 
and may hold the key for the cosmological constant problem
as discussed below. In the case of 2-dimensional 
NC spaces, \eq{covariant-rescaling} has 
no solution\footnote{I would like to 
thank A. Much for related discussions}, 
so that the
action cannot be written in standard form at all. This will
be discussed elsewhere.

The emergence of such noncommutative vacua
is very compelling in closely related (Euclidean) matrix 
models admitting compact NC branes as 
vacua \cite{Behr:2005wp,Grosse:2004wm}, 
and supported  by a considerable body of analytical 
and numerical work at least in 2 dimensions, including 
\cite{Azuma:2004qe,Azuma:2002zi,Azuma:2005pm,DelgadilloBlando:2008vi}
and references therein. In higher dimensions, it may be necessary
to consider supersymmetric matrix models
 as discussed below, cf. \cite{Azeyanagi:2008bk}.

\paragraph{Relation with string theory.}

In string theory, a somewhat related situation occurs in the context
of D-branes in  a nontrivial B-field background. This leads 
to an effective description
in terms of NC Yang-Mills theory on a 
noncommutative D-brane with Poisson structure $\theta^{\mu\nu}$
inherited from the B field, see e.g. \cite{Seiberg:1999vs} and references
therein. This effective gauge theory is governed
by the open string metric  \cite{Seiberg:1999vs}
which is strongly reminiscent of 
$\tilde G^{\mu\nu}(x)$ (apart from
the density factor), while $g_{\mu\nu}(x)$ 
corresponds to the  closed string metric (more precisely its pull-back on the brane).
Most of these results are restricted to the case of 
constant $\theta^{\mu\nu}$ and slowly varying fields, while the case of general
NC curved branes has received only limited attention, 
notably \cite{Cattaneo:1999fm,Cornalba:2001sm}.

However,
the results of the present paper should be compared more
properly with previous work on 
string-theoretical matrix models such as the 
IKKT model \cite{Ishibashi:1996xs}. NC branes have indeed been studied
in considerable detail in this context, 
and it is well-known that the matrix models can be interpreted as NC
gauge theory on the brane.
However, this has been worked out only for NC branes with a
high degree of symmetry, such as fuzzy spaces 
(see e.g. \cite{Chepelev:1997ug,Kabat:1997im,Myers:1999ps,
Alekseev:2000fd,Nair:1998bp,Azuma:2004qe}), 
or other special branes satisfying a 
BPS condition \cite{Banks:1996nn,Cornalba:1998zy,Aoki:1999vr,Taylor:2001vb}.
The role of the
effective metric \eq{G-def-general} is well-known in these cases, and evidence
for the existence of gravitons on the branes has been obtained
\cite{Kitazawa:2006pj}. For generic NC branes in matrix models,
the effective metric $\tilde G^{\mu\nu}(x)$ and its role 
in the effective field theory on
branes has not been elaborated previously, 
to the best knowledge of the author. 
Moreover, it is essential to note that the would-be $U(1)$
gauge field on the brane is absorbed in $\tilde G^{\mu\nu}(x)$, 
leading
to a dynamical emergent gravity. Therefore the present approach
could be seen as a novel way of obtaining
gravity from string-theoretical matrix models, avoiding the
conventional picture of string compactification.

In this context, it is worth recalling the relation between 
the semi-classical action \eq{S-semiclassical-general} and the Dirac-Born-Infeld
action for $\theta^{-1}_{\mu\nu} := B_{\mu\nu} + F_{\mu\nu}$
which governs the dynamics of branes in string theory \cite{Seiberg:1999vs}.
The action \eq{S-simple} arises from the DBI action at 
leading ``nontrivial'' order,
\be
\sqrt{\det(g_{\mu\nu} + \theta^{-1}_{\mu\nu})} \, \sim \,
\rho(x)\, (1 + 2\, \eta(x) + ...) 
\ee
omitting all constants, cf. \cite{Cornalba:1999ah}.

\paragraph{Equation of motion for test particle $\varphi$.}

The covariant e.o.m. for $\varphi$  obtained from 
the semi-classical action
\eq{covariant-action-scalar} is
\be
\Delta_{\tilde G} \varphi 
= (\tilde G^{\mu\nu}\partial_{\mu}\partial_{\nu} 
-\tilde\Gamma^\mu \partial_\mu)
\varphi = 0 \,.
\label{eom-phi-covar}
\ee
On the other hand, starting from the matrix model \eq{covariant-action-scalar-0}
we obtain the e.o.m. for the same scalar field $\varphi$ as 
\bea
0 = [X^a, [X^{b},\varphi]] \eta_{ab} &=& 
[X^\mu, [X^{\nu},\varphi]]\eta_{\mu\nu} 
 + [\phi^i, [\phi^j,\varphi]]\,\d_{ij} \nn\\
&=&  i[X^\mu, \theta^{\nu\eta}\partial_\eta\varphi] \eta_{\mu\nu} 
 + i[\phi^i,\theta^{\mu\nu}\partial_\mu\phi^j\partial_\nu\varphi]\,\d_{ij} \nn\\
&=&  -\theta^{\mu\rho}\partial_\rho
  (\theta^{\nu\eta}\partial_\eta\varphi) \eta_{\mu\nu} 
 -\theta^{\rho\sigma}\partial_\rho \phi^i
 \partial_\sigma(\theta^{\mu\nu}\partial_\mu\phi^j\partial_\nu\varphi) \,\d_{ij}
\nn\\
&=&  -\(\eta_{\mu\nu} \theta^{\mu\rho}\partial_\rho \theta^{\nu\eta} 
 +\theta^{\rho\sigma}\partial_\rho \phi^i
 \partial_\sigma(\theta^{\mu\eta}\partial_\mu\phi^j)\,\d_{ij}
 \)\partial_\eta\varphi  \nn\\
&&-\theta^{\mu\rho} \theta^{\nu\eta}
 (\eta_{\mu\nu} + \d g_{\mu\nu}) \partial_\rho\partial_\eta\varphi\nn\\
&\stackrel{\rm e.o.m.}{=}&  - G^{\rho\eta} \partial_\rho\partial_\eta\varphi \,.
\label{eom-extradim-phi}
\eea
The last equality holds for on-shell geometries
defined by \eq{eom-extradim}, and 
$\d g_{\mu\nu} \equiv \partial_\mu\phi^i \partial_\nu\phi^j\,\d_{ij}$. 
Comparing with the covariant form \eq{eom-phi-covar}, it follows that
\be
 \tilde G^{\mu\nu} \partial_\mu\partial_\nu\varphi = 0 
= \Delta_{\tilde G} \varphi, 
\label{eom-varphi}
\ee
for on-shell geometries, which implies the ``harmonic gauge''
\be
\tilde\Gamma^\mu \,\,\stackrel{\rm e.o.m.}{=} 0  .
\label{gauge-fixing}
\ee
This holds only in the preferred $x^\mu$ coordinates 
defined by the matrix model; a direct derivation 
based on \eq{eom-extradim} is given in Appendix A.

\paragraph{Equation of motion for  $X^a$.}

The same argument as above gives the equations of motion 
for the embedding functions $\phi^i$ in the matrix model \eq{YM-action-extra}, 
\be\fbox{$
\Delta_{\tilde G} \phi^i =0   $}
\label{eom-phi}
\ee
and similarly for $x^\mu \sim X^\mu$,
\be 
\Delta_{\tilde G} x^\mu = 0 .
\label{eom-X-harmonic-tree}
\ee
This reflects the freedom of choosing the separation of 
$X^a = (X^\mu,\phi^i)$
into coordinates and scalar fields.
In particular, on-shell geometries \eq{eom-extradim}
imply harmonic coordinates, which in
General Relativity \cite{weinberg}
would be interpreted as gauge condition.
We will now derive an equivalent but 
more useful form of \eq{eom-X-harmonic-tree} in terms of the 
``tangential'' $\theta^{-1}_{\mu\nu}(x)$:

\paragraph{Equation of motion for  $\theta^{-1}_{\mu\nu}(x)$.}

Reconsider the e.o.m. for the tangential components $X^\mu$
from the matrix model \eq{YM-action-extra}:
\bea
0 = [X^b, [X^{\nu},X^{b'}]] \eta_{bb'} &=& 
[X^\mu, [X^{\nu},X^{\mu'}]] \eta_{\mu\mu'} 
 + [\phi^i, [X^{\nu},\phi^j]]\,\d_{ij} \nn\\
&=& -\theta^{\mu\rho}\partial_\rho\theta^{\nu\mu'} \eta_{\mu\mu'} 
  -\theta^{\mu\rho}\partial_\mu\phi^i
\partial_\rho(\theta^{\nu\eta}\partial_\eta \phi^j \,\d_{ij}) \nn\\
&=& -\theta^{\mu\rho}\partial_\rho\theta^{\nu\eta} 
 (\eta_{\mu\eta} + \d g_{\mu\eta})
 -\theta^{\nu\eta}\theta^{\mu\rho}\partial_\rho \d g_{\mu\eta} \nn\\
&=& -\theta^{\nu\nu'} G^{\rho\eta'}(x) 
  \partial_\rho\theta^{-1}_{\nu'\eta'} 
 - \theta^{\nu\eta}\theta^{\mu\rho}\partial_\rho g_{\mu\eta} 
\label{eom-extradim}
\eea
since 
$\partial_\rho \d g_{\mu\eta}(x) = \partial_\rho g_{\mu\eta}(x)$, 
i.e.
\be
 G^{\rho\eta}(x) \partial_\rho\theta^{-1}_{\eta\nu} 
= \theta^{\mu\rho}\partial_\rho g_{\mu\nu}(x) \equiv J_\nu .
\label{eom-extradim-theta}
\ee
These are essentially Maxwell equations coupled to an external
current $J_\nu$, which depends on the matter field $\phi$.
As shown in Appendix B, 
this can be written in covariant form as
\be
\fbox{$
\tilde G^{\g \eta}(x)\, \tilde\nabla_\g (e^{\sigma} \theta^{-1}_{\eta\nu}) 
\, = \, e^{-\sigma}\,\tilde G_{\mu\nu}\,\theta^{\mu\g}\,\partial_\g\eta(x)$}
\label{eom-geom-covar-extra}
\ee
Here $\tilde \nabla$ denotes the Levi-Civita connection with respect to
the effective metric $\tilde G^{\mu\nu}$ \eq{G-tilde-general}, which is
no longer unimodular in general.
This has the same form as \eq{eom-geom-covar-1},
and  can be rewritten  
as 
\bea
\tilde G^{\g \eta}(x)\, \tilde\nabla_\g \theta^{-1}_{\eta \nu} 
 &=& - \tilde G^{\g \eta}(x)\, \theta^{-1}_{\eta \nu}
\partial_\g \sigma + 
e^{-2\sigma}\,\tilde G_{\mu\nu} \theta^{\mu\g}\,\partial_\g\eta(x)
\label{eom-covar-2}
\eea
The derivation in Appendix B assumes that the embedding 
functions $\phi^i$ also satisfy their e.o.m.
\eq{eom-phi}. It can also be derived directly from the 
semi-classical action \eq{S-semiclassical-general}:

\paragraph{Semi-classical  derivation of e.o.m. for  $\theta^{-1}_{\mu\nu}(x)$.}

Starting from \eq{S-semiclassical-general}, we can derive the covariant
e.o.m. of the matrix model using 
\be
\d \theta^{-1}_{\mu\nu} = 
\tilde \nabla_\mu \d A_\nu - \tilde \nabla_\nu \d A_\mu .
\ee
This gives
\bea
\d S_{YM} &=& 2\int  d^{2n}x\, \Big(\d \eta(x)\, 
\sqrt{\det\theta^{-1}_{\mu\nu}} + \eta(x)\, 
\frac 1{2\sqrt{\det\theta^{-1}_{\mu\nu}}}\, \det\theta^{-1}_{\mu\nu}
(\theta^{\mu\nu}\d \theta^{-1}_{\nu\mu}) \Big)\nn\\
&=& \int  d^{2n}x\, \rho
\Big(g_{\mu\nu} \theta^{\mu\mu'}\d\theta^{\nu\nu'}g_{\mu'\nu'} \,  
+ g_{\mu\nu} \theta^{\mu\mu'}\theta^{\nu\nu'}\d g_{\mu'\nu'} \, 
+ \,\eta(x)\,(\theta^{\mu\nu}\d \theta^{-1}_{\nu\mu}) \Big)\nn\\
&=& \int  d^{2n}x\,\rho
\Big(G^{\eta\mu}\theta^{-1}_{\mu\nu}G^{\nu\rho}
\d\theta^{-1}_{\rho\eta}\,
+ G^{\mu\nu} \d g_{\mu\nu}
+ \,\eta(x)\,(\theta^{\mu\nu}\d \theta^{-1}_{\nu\mu}) \Big)\nn\\
&=& 2\int  d^{2n}x\,\sqrt{|\tilde G|} 
\Big(\tilde G^{\eta\mu}\tilde G^{\nu\rho} 
e^{\sigma}\theta^{-1}_{\mu\nu}\tilde \nabla_\rho \d A_\eta\,
- \,e^{-\sigma}\eta\,\theta^{\rho\eta} 
\tilde \nabla_\rho \d A_\eta\, \Big) 
\nn\\
&& \qquad +2\int  d^{2n}x\,
 \sqrt{|\tilde G|}\, \tilde G^{\mu\nu} \partial_\mu \phi^i  \partial_\nu \d\phi^i\, \d_{ij}
\eea
using 
\be
\rho = |\tilde G_{\mu\nu}|^{1/2}\, e^{-\sigma} 
\label{rho-a-id}
\ee 
which follows from \eq{covariant-rescaling}. Noting that 
\be 
\int d^{2n}x\, \sqrt{|\tilde G|}\, \tilde \nabla_\mu V^\mu =0
\ee
and $\tilde \nabla \tilde G =0$ 
we obtain
\bea
\d S &=& -2\int  d^{2n}x \, \sqrt{|\tilde G|} \,\d A_\eta 
\Big(\tilde G^{\eta\mu}\tilde G^{\nu\rho}
\tilde \nabla_\rho(e^{\sigma} \theta^{-1}_{\mu\nu})\,
- \,\tilde \nabla_\rho(e^{-\sigma}\eta\,\theta^{\rho\eta}) \Big)
 + \d\phi^i\,\d_{ij} \partial_\nu 
\(\sqrt{|\tilde G|}\, 
\tilde G^{\mu\nu}\tilde \partial_\mu \phi^i \) \nn\\
&=& -2\int  d^{2n}x\, \sqrt{|\tilde G|} \,\(\d A_\eta 
\Big(\tilde G^{\eta\mu}\tilde G^{\nu\rho}
\tilde \nabla_\rho(e^{\sigma} \theta^{-1}_{\mu\nu})\,
-\,|\tilde G|^{-1/2} \partial_\rho 
(|\tilde G|^{1/2}e^{-\sigma}\eta \theta^{\rho\eta})\Big) 
 + \d\phi^i\,\d_{ij} \Delta_{\tilde G} \phi^i \) \nn\\
&=& -2\int  d^{2n}x\,\sqrt{|\tilde G|} \,\(\d A_\eta 
\Big(\tilde G^{\eta\mu}\tilde G^{\nu\rho}
\tilde \nabla_\rho(e^{\sigma} \theta^{-1}_{\mu\nu})\,
-e^{-\sigma}\theta^{\rho\eta}\,\partial_\rho \eta \Big)
 + \d\phi^i\,\d_{ij} \Delta_{\tilde G} \phi^i\)\, \nn
\label{variation-bare-geometry}
\eea
using \eq{partial-theta-id} and \eq{rho-a-id} in the last steps.
This gives precisely the equations of motion 
\eq{eom-geom-covar-extra} and \eq{eom-phi}.

\paragraph{Formal considerations.}

From a more formal point of view, we have the following 
structures: The submanifold $\cM^{2n} \subset \R^D$
carries an embedding metric $g$, and a preferred
frame $e^\mu = \theta^{\mu\nu}\partial_\nu \sim [X^\mu,.]$ 
which encodes the noncommutative structure. 
The effective metric $G^{\mu\nu}$ 
\eq{G-def-general} on $T^* \cM$ is defined by
\be
(\b,\b')_G := (\star \b,\star \b')_g, \qquad \b,\b' \in T^*\cM \, ,
\ee
where $(\partial_\mu,\partial_\nu)_g = g_{\mu\nu}(x)$ and
 $\star: T^*\cM \to T\cM$ is the canonical map defined by
the Poisson structure $\theta^{\mu\nu}$.

Notice the unusual role of the indices. This makes sense here,
because the frame 
$e^\mu$ is given in terms of the antisymmetric Poisson structure 
$\theta^{\mu\nu}$ in the preferred coordinates $x^\mu$. 
There is no distinction between ``Lorentz'' and ``covariant'' indices
here, and neither local Lorentz nor general coordinate
transformations are allowed a priori.
One could proceed to introduce differential forms in terms of 
one-forms $\theta^a,\,\,a=1,..., D$ through
$[\theta^a,X^b] =0,\,\, \theta^a \theta^b = - \theta^b \theta^a$.
The exterior differential of functions is then defined 
in terms of a ``special'' one-form $\theta$,
\be
df = [\theta,f] , \qquad
\theta = X^a \eta_{ab} \theta^b .
\ee
This is similar\footnote{However the frame and metric here 
have a specific form in terms of $\theta^{\mu\nu}$, 
unlike in \cite{Madore:2000aq}.} 
to the formalism in \cite{Madore:2000aq,Madore-book}; however the calculus is 
$D$-dimensional, similar to the case of 
the fuzzy sphere \cite{Madore:1991bw}. 
The scalar action can then be written as 
\be
S[\varphi] \sim \int  d^{2n} x\, \rho\, \langle d\phi, d\phi\rangle
\ee
where $\langle\theta^a,\theta^b\rangle = \eta^{ab}$. Similarly, the
semi-classical form of the matrix model is 
\be
S_{YM} \sim \frac 1{(2\pi)^n}\,
\int  d^{2n} x\, \rho\, \langle \theta\wedge\theta,
\theta\wedge\theta\rangle.
\ee 
We can also write
\bea
df &=& [\theta,f] \, = e^\mu(f)\, \tilde\theta_\mu , \nn\\
\tilde\theta_\mu &=& \eta_{\mu\nu} \theta^\nu + \partial_\mu \phi^i
\delta_{ij} \theta^{2n+j}
\eea
where $\tilde\theta_\mu$ is in some sense dual to $e^\mu$.
This should illuminate the relation and difference to 
\cite{Madore:2000aq}.
These considerations will be pursued further elsewhere.

\subsection{Nonabelian gauge fields}

Now consider backgrounds of the form
\be
Y^a = X^a \otimes \one_n + \cA^a_\a \otimes \lambda^\a
\ee
where $\la^\a$ are generators of $su(n)$. According to
\cite{Steinacker:2007dq}, the $U(1)$ sector  
(i.e. the components
proportional to $\one_n$) is absorbed in the
geometrical degrees of freedom defined by $X^a$,
and the discussion of the previous sections applies without
change. On the other hand, the $su(n)$ components
$\cA^\mu_\a$ behave as nonabelian gauge fields, and similarly
the transversal $su(n)$ components $\phi^i_\a$ in
\be
\phi^i =  \bar\phi^i \otimes \one_n + \phi^i_\a \otimes \la^\a
\ee
are nonabelian scalars from the brane point of view. The 
$\phi^i_\a$ then propagate in the background geometry $\tilde
G_{\mu\nu}$ as discussed above.
If some of the $\phi^i_\a$ develop a nontrivial vev, they might
be viewed as part of the geometry.

It was  shown in \cite{Steinacker:2007dq} that
the effective action for nonabelian 
gauge fields $\cA^\mu_\a$ due to the 4-dimensional matrix model 
\eq{YM-action-1}
in the semi-classical limit is
\bea
S_{YM}[\cA] &\sim&  \int d^4 x\, \rho(x) 
tr \(G^{\mu\mu'} G^{\nu\nu'} F_{\mu\nu}\, F_{\mu'\nu'}\)
 + 2  \int \eta(x) \, tr F\wedge F  \nn\\
&=&  \int d^4 x\, |\tilde G_{\mu\nu}|^{1/2}
e^{\sigma}\,\tilde G^{\mu\mu'} \tilde G^{\nu\nu'}
 tr F_{\mu\nu}\, F_{\mu'\nu'}
 + 2  \int \eta(x) \, tr F\wedge F  \; .
\label{S-YM-effective-nonabelian}
\eea
This is the Yang-Mills action for a nonabelian gauge fields coupled
to the effective metric $\tilde G_{\mu\nu}$, 
apart from the ``would-be topological term'' and 
the density factor $e^{\sigma}$. The latter could be interpreted
as varying bare gauge coupling ``constant'' 
\be
g_{YM}^2 = g^2 e^{-\sigma}
\label{YM-coupling-effective}
\ee
introducing an overall coupling constant $\frac 1{g^2}$ 
to the matrix model \eq{YM-action-1}.
In order to be  physically acceptable, it is probably required that 
$\sigma$ is slowly varying.
Indeed, a kinetic term for the ``dilaton'' $\rho$
resp. $\sigma$ is induced in the quantum effective action
\cite{Klammer:2008df},
except in the case of unbroken $N=4$ supersymmetry.
This might ensure that $\sigma$ is nearly constant.

Due to the strong constraints of gauge invariance,
we expect that \eq{S-YM-effective-nonabelian} applies
without change to the case of non-trivially
embedded 4-dimensional branes in $\R^D$; 
however this remains to be shown. 
Note that $\eta \sim e^{\sigma}$ due to \eq{eta-scale}, hence
the two terms in \eq{S-YM-effective-nonabelian} have roughly
the same coefficients.
This changes for higher-dimensional branes, where
the ``would-be topological term'' 
$\int \eta(x) \, tr F\wedge F$
will be replaced by a different term which could be determined 
along the lines in \cite{Steinacker:2007dq}.
 Before relating this e.g. to the strong CP problem
one would first have to identify more realistic models, elaborate 
the symmetry breaking etc..

\subsection{Fermions}

Then the most obvious 
(perhaps the only reasonable) action
for a spinor which can be written down in the matrix 
model framework\footnote{In particular, 
fermions should also be in the adjoint, otherwise they
cannot acquire a kinetic term. 
This does not rule out its applicability
in particle physics, see e.g. \cite{Steinacker:2007ay}.} is
\bea
S &=& (2\pi)^{2n}\, \Tr\, \obar\Psi  \gamma_a [X^a,\Psi]
\sim \,\, \int d^{2n} x\, \rho(x)\, \obar\Psi i 
(\gamma_\mu + \gamma_{2n+i}\partial_\mu \phi^i)
 \theta^{\mu\nu}(x)\partial_\nu\Psi   \nn\\
&=& \,\, \int d^{2n} x\, \rho(x)\, \obar\Psi i 
\tilde \gamma_\mu
 \theta^{\mu\nu}(x)\partial_\nu\Psi  
\label{fermionic-action-geom}
\eea
where $\gamma_a$ defines the D-dimensional Euclidean Clifford
algebra, and
\be
\tilde \gamma_\mu = \gamma_\mu  + \gamma_{2n+i} \partial_\mu \phi^i
\ee
satisfies the Clifford algebra associated with the embedding 
metric $g_{\mu\nu}(x)$ on $\cM$,
\be
\{\tilde \gamma_\mu, \tilde \gamma_\nu\} 
= 2 \eta_{\mu\nu} 
+ 2 \partial_\mu \phi^i\, \partial_\nu \phi^j \d_{ij} 
= 2 g_{\mu\nu}(x) \, . 
\ee
This is indeed the appropriate coupling of a spinor to the 
background geometry with metric $G_{\mu\nu}$ (up to rescaling),
albeit with a non-standard spin connection 
which vanishes in the $x^\mu$ coordinates. 
This nicely generalizes at the classical level 
the analysis in \cite{Klammer:2008df}, where
this action was shown to
provide a reasonable coupling of fermions to emergent 
gravity for flat $g_{\mu\nu}$.
At the quantum level, it was shown in \cite{Klammer:2008df}
that the Einstein-Hilbert term is indeed induced (along 
with a Dilaton-like term for $\sigma$), 
for flat $g_{\mu\nu}$ and on-shell 
geometries. It remains to be verified whether this generalizes
to the case of non-trivially embedded branes.
This is expected to be the case since it does give 
the correct Dirac operator e.g. in the case
of the fuzzy sphere \cite{Grosse:1994ed} or for 
$S^2_N \times S^2_N$ \cite{Behr:2005wp}.

Given the above Dirac operator, one could also consider the 
associated spectral action 
in the sense of \cite{Chamseddine:1996zu}.
It is an open problem in that context how to quantize gravity,
more precisely how to integrate over the various geometries. 
The present framework suggests a simple answer: 
The Dirac operator should have the form as given in \eq{fermionic-action-geom}, and
the integral over the geometries should be realized as integral over the 
matrices $X^a$ {\em with measure defined by the bosonic matrix model},
$d\mu(X^a) = e^{-S_{YM}[X^a]}$ \eq{YM-action-extra}. 
Nevertheless, this is not entirely
equivalent to the present matrix model framework:  
The spectral action is based on the dependence of the spectrum as a
function of the cutoff, while in the 
$N=4$ case as considered here such a cutoff should not be required.

\subsection{Compactification of branes}
\label{sec:compactification}

Consider a $2n$--dimensional NC brane
$\cM_\theta^{2n}\subset \R^{10}$.
In order to obtain a 4-dimensional space at low energies, 
we assume that this
higher-dimensional brane has compact extra dimensions,
for example
\be
\cM^{2n}_\theta \sim \cM^4_\theta \times K_\theta \, .
\label{NC-compactification}
\ee
If $K$ is ``small'' enough, this looks like 
$\cM^4_\theta$ at low energies, as in standard compactification
scenarios.
Particularly natural examples would be 
$\cM^6_\theta \sim M^4_\theta \times S^2_N$ or 
$\cM^8_\theta \sim M^4_\theta \times S^2_N\times S^2_{N'}$ where
$S^2_N$ 
denotes the fuzzy sphere. Such extra-dimensional fuzzy spaces can
indeed be embedded naturally in the matrix models considered here
(possibly upon adding soft SUSY-breaking terms) 
\cite{Azuma:2005pm,Azuma:2002zi}, or
alternatively they can arise spontaneously from the scalar fields
from the 4-dimensional point of view
\cite{Aschieri:2006uw,Steinacker:2007ay}. These 2 points
of view are essentially equivalent.

Let us count degrees of freedom for the effective metric. 
For a $2n$ - dimensional NC brane, $\theta^{\mu\nu}$ 
resp. $\theta^{-1} = dA$ has 
$2n-2$ physical (on-shell) plus one off-shell degrees of freedom,
after gauge fixing. Upon compactification on $K_\theta$,
the components $A_i$  tangential to $K_\theta$ become massive,
leaving only 2 massless d.o.f. from a 4-dimensional 
point of view.
The embedding of $M^{2n} \subset \R^{10}$ defined by $\phi^i$
provides $10-2n$ additional degrees of freedom.
They are absorbed in the effective 
metric and governed by the quantum effective action.
From the 4-dimensional point of view, this will lead to an 
effective ``brane tension'' on $\cM^4_\theta$ depending
on the moduli of the compactification (e.g. the radius)
as indicated below. Those are likely to become ``off-shell'' d.o.f.
which enlarge the class of effective 4-dimensional metrics, as desired. 
Therefore one can expect to recover most of the 
2 on-shell plus 4 off-shell d.o.f. of the 4-dimensional metric
in General Relativity.
All this requires a more detailed analysis.

\paragraph{Example 1: The fuzzy sphere $S^2_N$.}

The fuzzy sphere $S^2_N$ \cite{Madore:1991bw}
is a natural realization of this framework, being realized in terms of an 
embedding $S^2 \subset \R^3$. Consider our matrix model
in $D=3$, with the configuration 
\bea
X^a &=& \frac{r}{c_N}\, \lambda^{(N),a}, \qquad a=1,2,3  \nn\\
\left[X^a,X^b\right] &=& i \theta_N \varepsilon^{abc}\, 
  \frac{X^{c}}{r}, \qquad
 X^{a} X^{a'}\,\d_{aa'} =  \frac{r^2}{c_N^2}\,\frac 14 (N^2-1)
 \, =\,  r^2 .
\label{fuzzy-sphere}
\eea
Here $r$ is an arbitrary radius, and
$\lambda^{(N),a}$ denotes the generators of the $N$-dimensional
irreducible representation of $SU(2)$, 
\bea
c_N^2 &=&  \frac 14 (N^2-1)      \label{c-N}\\
\theta_N &=& \frac{r^2}{c_N} . \label{theta-N}
\eea 
Even though \eq{fuzzy-sphere} is not a solution of the 
basic matrix model \eq{YM-action-1}, it makes nevertheless sense
to consider such configurations since the
induced gravity action is not yet taken into account.
Moreover, it becomes a solution once a 
mass term and/or a cubic term is added to the action, e.g.
\be
S_{\rm YM} + S_{\rm corr} = (2\pi) Tr \Big([X^a,X^b][X^{a'},X^{b'}]
\d_{aa'}\d_{bb'} + m^2\, X^a X^{a'}\d_{aa'} 
+ \c \, X^a X^b X^c \varepsilon_{abc} \Big) .
\label{S-fuzzy-additional}
\ee
Such terms might be induced in the quantum 
effective action, possibly after SSB.
The general discussion of Section \ref{sec:extradim}
applies as follows:
consider e.g. some neighborhood of the north pole 
$x^3 \approx r, \, x^1
\approx x^2 \ll r$ of $S^2$. Then we separate the 
coordinates as in \eq{extradim-splitting}
in 2 tangential ones and one scalar ``embedding'' function,
\be
X^a = (X^\mu,\phi), \qquad \mu = 1,2
\label{extradim-splitting-S2}
\ee
where $\phi = \phi(X^\mu) \approx 0$ for small $X^1,X^2$. 
Indeed it is not difficult to write $\phi = X^3$ in \eq{fuzzy-sphere}
as a function of $X^1,X^2$, in a suitable domain. 
Hence $S^2_N$ is a NC brane embedded in $\R^3$. 
The Poisson tensor e.g. at the north pole is
$\theta^{12} = \theta_N$, and $S^2_N$ is a quantization of 
$S^2$ with the symplectic structure
\be
\omega_{S^2} 
= \theta_N^{-1}\,\frac 1{r}\, \varepsilon_{abc}\, x^a\, dx^b dx^c ,
\ee
where $x^a$ is the semi-classical limit of $X^a$.
It satisfies the semi-classical quantization condition 
\be
2\pi N = \int_{S^2} \omega_{S^2} = \int_{S^2} d^2 x\, \rho  
= \theta_N^{-1}\, 4\pi r^2
  = 4\pi c_N
\label{integral-S2}
\ee
consistent with \eq{c-N}, where $\rho = \theta_N^{-1}$. 
Therefore $S^2_N$ can be considered
as a compactification of the $D=2$ Moyal-Weyl plane.
The embedding metric $g_{\mu\nu}(x)$ is the round metric for
a sphere $S^2$ with radius $r$, and $G^{\mu\nu} = \theta_N^2
g^{\mu\nu}$.

\paragraph{Example 2: $\R^4_\theta \times S^2_N$.}

Now consider 
a configuration $\cM^6_\theta = \R^4_\theta \times S^2_N$. 
This can be realized in  the $D$-dimensional matrix model 
for $D \geq 7$
\bea
X^\mu &=& \bar X^\mu, \qquad \mu = 0,1,2,3\, \nn\\\
\phi^i &=&  \frac{r}{c_N}\, \lambda^{(N),i}, \qquad i=1,2,3  
\eea
where $\bar X^\mu$ are the generators of $\R^4_\theta$ \eq{Moyal-Weyl}.
This should be interpreted as a 6-dimensional NC space, which 
for small $r$ looks like
$\R^4_\theta$. Such configurations can lead to 
interesting low-energy gauge groups and zero modes in the nonabelian 
case, as discussed in 
\cite{Aschieri:2006uw,Steinacker:2007ay}. 
Similar configurations were discussed previously
in the IKKT model \cite{Azuma:2005pm}, see also \cite{Aoki:2008qt}. 
The radius of the fuzzy spheres will be dynamical
$r = r(X^\mu) \sim r(x^\mu)$, determined by the 
effective action. 
Inserting this configuration in the action and recalling 
\eq{S-semiclassical-general}, we obtain
\bea
S_{YM} &=& (2\pi)^3\, Tr[X^a,X^b][X^{a'},X^{b'}] \d_{aa'} \d_{bb'}
\sim 4 \int_{\R^4\times S^2} d^4 x\, 
\omega_{S^2}\rho^{(4)}(x)\,\eta^{(6)}(x) \nn\\
&=& 8\pi N\,\int_{\R^4} d^4 x\, \rho^{(4)}(x)\, \eta^{(6)}(x)
\label{S-tree-S2}
\eea
using \eq{integral-S2} and
\be
\eta^{(6)}(x) \sim 
[X^a,X^b][X^{a'},X^{b'}]\d_{aa'} \d_{bb'} \sim \eta^{(4)}(x)
+ 2 G^{\mu\nu}\partial_\mu r(x) \partial_\nu r(x)  
+ 2 \frac{r(x)^4}{c_N^2} ,
\ee
cf. \eq{eta-def}. Here
\be
\eta^{(4)}(x) = G^{\mu\nu} g_{\mu\nu}(x)
\ee
involves only the 4-dimensional metric. This leads to an effective
potential  $V(r)$ for the radius $r(x)$, which will receive
additional contributions from further terms such as 
\eq{S-fuzzy-additional} and from the induced gravitational action.
For example, 
consider the 6-dimensional ``would-be cosmological
constant'' term in \eq{S-oneloop-gravity}, which using
\eq{G-tilde-general} can be written as 
\bea
\int  d^6 x\,\sqrt{|\tilde G|}\, \L^6 
&\sim& \int d^6 x\, |\theta^{-1}_{\mu\nu}|^{1/4}\,|g_{\mu\nu}|^{1/4} \L^6  \nn\\
 &=&  4\pi c_N^{1/2}\,\int d^4 x\,  r(x)\, 
\rho(x)^{1/2}\,|g_{\mu\nu}|^{1/4} \, \L^6 
\label{S-inducedL4-S2}
\eea
where $\rho(x)$ and $g_{\mu\nu}(x)$ in the last
line are 4-dimensional quantities.
Somewhat surprisingly, this term now depends on
$\rho(x)$, unlike in the case of 4D branes. 
However, this expression should be taken with much caution, 
because the IR condition
\eq{IR-regime} for the applicability of the semi-classical 
expressions \eq{S-oneloop-gravity}
may not be appropriate for the compact dimensions. 
In that case, it might be more appropriate to use a 
4-dimensional description rather than the 
6-dimensional metric.
In any case, this will contribute to the effective potential for
$V(r)$, but a more detailed analysis is required.

\paragraph{Example 3: $\R^4_\theta \times S^2_{N_L}\times S^2_{N_R}$.}

A generalization of the above configuration  which 
can be realized in the $10$-dimensional IKKT matrix model is
\bea
X^\mu &=& \,\bar X^\mu, \qquad \mu = 0,1,2,3\, \nn\\\
\phi^i &=& \frac{r_L}{c_{L}}\, \lambda^{(N_L),i}, \qquad i=1,2,3, 
\nn\\
\phi^i &=& \frac{r_R}{c_{R}}\, \lambda^{(N_R),i}, \qquad i=4,5,6  
\eea
which should be interpreted as a 8-dimensional NC space. 
The effective 4-dimensional action now involves 
2 parameters $r_L, r_R$, which will be governed by an effective 
potential $V(r_L,r_R)$. This should provide sufficient
structure to obtain interesting solutions from the particle
physics point of view; see also
e.g. \cite{Azuma:2005pm,Behr:2005wp} and references therein for 
related work.

\subsection{Departures from General Relativity: 
preferred scales and coordinates} 
\label{sec:departures}

There are several features of the model under consideration which 
differ radically from the conventional picture of General Relativity. 
We focus on the case of 4-dimensional NC branes for simplicity.

First, recall that there are preferred coordinates
in the model, given by the covariant coordinates $x^\mu$. 
In those coordinates, the background metric 
is explicitly constant, $g_{\mu\nu} = \d_{\mu\nu}$ resp.
$g_{\mu\nu} = \eta_{\mu\nu}$ in the $D=4$ case,
and the preferred frame is given by the antisymmetric Poisson tensor,
$e^\mu = \theta^{\mu\nu}\, \partial_\nu$. 
This is physically not very significant a priori, but it 
simplifies the issue of gauge fixing. We recall that 
in the preferred $x^\mu$ coordinates, the on-shell condition
for $X^\mu$ amounts to \eq{gauge-fixing}
\be
\tilde\Gamma^\mu = 0\, ,
\ee
which would be interpreted as gauge choice
in General Relativity. 

A more significant feature of the matrix model
is the presence of the scalar density 
$\rho = (\det \theta^{-1}_{\mu\nu})^{1/2}$ \eq{rho-def-general}, 
which defines the 
scale of noncommutativity
\be
\rho  = \L_{NC}^4 =  L_{NC}^{-4}
\ee
and provides the symplectic measure 
$(2\pi)^2 \Tr f \sim \int d x\, \rho\, f$. 
Such a structure does not exist in
the commutative framework.  This leads to an analog of
the Bohr-Sommerfeld quantization condition, 
\be
{\rm Vol}_\theta =  (2\pi)^2 \cN
\label{Vol-x-relation}
\ee
where ${{\rm Vol}_\theta}$ denotes the volume measured 
in units of $L_{NC}$, and $\cN$ the dimension of the corresponding
Hilbert (sub)space. This means that the volume 
is quantized in integer multiples of $L_{NC}^4$, 
so that NC branes are automatically ``large'' for large
$\cN$. This is already a hint that NC spaces like to be flat, 
which is very interesting in connection with the cosmological
constant problem.

There is another scale in the model determined by the 
embedding metric $g_{\mu\nu}$ resp. the effective metric
$\tilde G_{\mu\nu}$, 
\be
L_g^4 = \L_{g}^{-4} = |\tilde G^{\mu\nu}|^{1/2}
\ee
which we could set to 1 thereby fixing the units;
recall that $|\tilde G_{\mu\nu}| \equiv |g_{\mu\nu}|$ 
\eq{G-g-4D} for general 4-dimensional branes in $\R^D$.
The ratio of these scales defines the dimensionless scalar function
\be
e^{-\sigma}
= \frac{|\theta_{\mu\nu}|^{1/2}}{|\tilde G_{\mu\nu}|^{1/2}}
= \frac{\L_{NC}^{4}}{\L_{g}^{4}}
\label{G-g-ratio}
\ee
using \eq{G-tilde-general}. 
We can relate this with the  Riemannian volume 
of $(\cM^4_\theta,\tilde G_{\mu\nu})$ measured by $\tilde G_{\mu\nu}$,
\be
{\rm Vol}_{\tilde G}
= (2\pi)^2\cN\, e^{\sigma} \, .
\label{Vol-x-G-relation}
\ee
The ``dilaton'' $e^{\sigma}$ will be determined dynamically 
by the model resp. the background
under consideration.
For example, in matrix models for fuzzy spheres it depends on the
coefficient of additional (soft SUSY-breaking)
terms such as $\Tr\, \varepsilon_{abc} X^a X^b X^c$, see also 
the related discussion in \cite{Aschieri:2006uw}.
Note that $e^{-\sigma}$ also gives the scale of $\eta$, 
\be
\eta \,\approx\, \frac{|\tilde
  G_{\mu\nu}|^{1/2}}{|\theta_{\mu\nu}|^{1/2}} 
\,= \, e^{\sigma} \, ,
\label{eta-scale}
\ee
at least for simple 4-dimensional configurations.
This may be a significant large dimensionless number.

In the context of quantization, 
we will encounter 2 additional scales 
$\L_4 \gg \L_1$ in the 10-dimensional version of the model,
where $\L_4$ 
is the scale of $N=4$ SUSY breaking which is argued to coincide
with the Planck scale $\L_{\rm Pl}$ below, and $\L_1$
is the scale of $N=1$ SUSY breaking. These should also be 
dynamical scales. We will furthermore argue that
$\L_{NC} > \L_4$ simplifies the 
semi-classical analysis, however this is  not essential;
it seems actually plausible that $N=4$ SUSY 
is broken by the NC background, so that $\L_{NC} = \L_4$.
In summary, we expect 3 a priori distinct physical scales
\be
\L_{NC} \geq \L_{4} = l_{\rm Pl}^{-1}  \gg \L_1 
\label{3-scales}
\ee
in addition to the dimensionless number $e^{\sigma}$ in the model.

\subsection{Quantization and induced gravity}
\label{sec:induced-grav}

Now consider the quantization of our matrix model,
which can contain scalar fields 
(such as e.g. arising from extra dimensions), fermions, 
the ``would-be $U(1)$ gauge field'' which is absorbed in 
$\theta^{\mu\nu}(x)$, and possibly nonabelian gauge fields.
In principle, the quantization is defined in terms of an 
integral
over all matrices. This is expected to be well-defined at least
in the case of the IKKT model, which leads to $N=4$ SUSY on $\R^4_\theta$
 \cite{Ishibashi:1996xs,Aoki:1999vr}.
Some modifications such as soft SUSY breaking terms may also be 
allowed.
Note that this quantization implies an integration over all
geometries of the NC branes embedded in $\R^{10}$, via the quantization
of the embedding functions $\phi^i$ as well as $X^\mu$
resp. $\theta^{\mu\nu}(x)$. In particular, (emergent) gravity is also quantized.

To obtain a qualitative understanding of the model at the quantum
level, we can take advantage of the above semi-classical form of the action 
in terms of conventional field theory coupled to $\tilde G^{\mu\nu}$.
Then the low-energy effective action at one loop 
can be extracted from
standard results of ordinary quantum field theory on
curves spaces. 
As shown in \cite{Grosse:2008xr,Klammer:2008df}, this is indeed
justified (based on a comparison with a fully 
NC computation and UV/IR mixing)
provided there exists an effective
UV-cutoff $\L$, and the following IR regime \cite{Grosse:2008xr}
is respected
\be
p\, \L < \L_{NC}^2 \, \quad\mbox{and}\quad \L < \L_{NC} \, .
\label{IR-regime}
\ee
These conditions ensure that the effects of noncommutativity are mild
even in the loops, so that the phase factors in non-planar diagrams
are small and are well approximated be the Poisson structure.
This reflects the fact that emergent NC gravity
is an IR phenomenon. A violation of e.g.
$\L < \L_{NC}$ is acceptable, but implies corrections\footnote{
if this condition is violated, a more refined analysis of NC 
corrections is required, cf. \cite{Grosse:2008xr}.
It turns out that the apparently-quartic divergent term
$\int \tilde\L_1^4 \sqrt{\tilde G}$ actually becomes milder,
being a difference between quadratically-divergent 
planar and non-planar diagrams. A similar comment applies to the 
$\int \L_4^2 R[\tilde G]$ term.} to the 
effective action \eq{S-oneloop-gravity} given below, 
some of which have been discussed in \cite{Grosse:2008xr}.
 Such a cutoff
is realized in the $N=4$ supersymmetric version of the 
model, assuming that $N=4$ SUSY is broken 
at $\L= \L_4$ from now on. This is essential, because  no
bare term in the action is available
which could cancel the induced (gravitational)
action discussed below.
We will furthermore assume that some smaller supersymmetry survives
down to a much lower energy scale $\L_1$, below which no 
supersymmetry survives. These are reasonable assumptions,
which appear to be necessary for the proposed framework
to be physically viable. 
Note that these scales are measured using
the physical metric $\tilde G_{\mu\nu}$.

The results of the one-loop computation of fields coupled to 
the background metric $\tilde G_{\mu\nu}$ can be obtained 
conveniently using the Seeley- de Witt coefficients
of the corresponding heat kernel.
The essential features are illustrated 
by the quantization of scalar fields.
Hence consider the
effective action obtained by integrating out the scalars, which in the
Euclidean case is 
\be
e^{-\Gamma_\phi[\tilde G]} = \int d\Phi\, e^{-S[\Phi]} \, .
\label{one-loop-action}
\ee
Since we are mainly interested in the induced gravitational action
here, it is sufficient to consider the case of non-interacting 
scalar fields coupled to the 
metric $\tilde G_{\mu\nu}$, where 
\be
\Gamma_\phi[\tilde G] = \frac 12 \Tr \log \frac 12\Delta_{\tilde G} \, 
\label{trlog}
\ee
assuming Euclidean signature for simplicity.
Here $\Delta_{\tilde G}$ is the Laplacian of a scalar field on
 the  Riemannian manifold $(\cM,\tilde G^{ab}(y))$
with action \eq{scalar-action-geom}. The UV cutoff 
$\L$ is incorporated using the Schwinger parametrization
\bea
\Tr \Big(\log\frac 12\Delta_{\tilde G}  - \log\frac 12\Delta_0\Big)
&\sim& -\Tr\int_{0}^{\infty} \frac{d\a}{\a}\,
(e^{-\a\frac 12\Delta_{\tilde G}} - e^{-\a\frac 12\Delta_0})\,\, \nn\\
&\equiv&\,\, -\Tr\int_{0}^{\infty} \frac{d\a}{\a}\,
\Big(e^{-\a\frac 12\Delta_{\tilde G}} - e^{-\a\frac 12\Delta_0
}\Big)\, 
e^{- \frac 1{\a\L^2}}\, .
\label{TrLog-id}
\eea
Now we can use the heat kernel expansion, 
\be
\Tr e^{-\frac 12\a\Delta_{\tilde G}} \sim \sum_{m \geq 0}\,(\frac{\a}2)^{m-n}
\int_{\cM}\,  d^{2n} x\,\sqrt{|\tilde G_{\mu\nu}|}\,\,
 a_{2m}(x,\Delta_{\tilde G})\, .
\label{Seeley-deWitt}
\ee
The $a_m(x,\Delta_{\tilde G})$ are known as Seeley-de Witt (or Duhamel)
coefficients, which for scalar fields with 
action \eq{scalar-action-geom} are given by \cite{Gilkey:1995mj} 
\bea
a_0(x) &=& \frac 1{(4\pi)^n}\, , \nn\\
a_2(x) &=& \frac 1{(4\pi)^n}\, \Big(\frac 16 R[\tilde G] \Big), \nn\\
a_4(x) &=& \frac 1{(4\pi)^n}\, \frac 1{360}\,  \(12 {{R_{;\mu}}}^\mu +
5 R^2 - 2 R_{\mu\nu} R^{\mu\nu} + 2
R_{\mu\nu\rho\sigma} R^{\mu\nu\rho\sigma}\) .
\label{SdW-coeff}
\eea  
The full effective action is complicated by the fact that 
there are several types of fields in the model including
scalars, fermions, and gauge fields. Because they all couple to the 
same effective metric up to possibly density factors\footnote{This was
shown in \cite{Steinacker:2007dq} for gauge fields and in 
\cite{Klammer:2008df} for fermions. 
Similar results are expected for the gravitons 
(i.e. the would-be $U(1)$ gauge field) due to
supersymmetry, at least in the case of $N=4$ SUSY.},
they will all induce essentially the same type of gravitational action,
with an additional kinetic term for the ``dilaton'' $\sigma$ 
due to the fermions \cite{Klammer:2008df} and gauge fields
resp. gravitons. Therefore one obtains
the following type of induced gravitational action:
\be 
\Gamma_{1-loop}[\tilde G] = \frac 1{(4\pi)^n}\, \int d^{2n} x \,
\sqrt{|\tilde G_{\mu\nu}|}\,\( -c_0\,\L_1^{2n} 
- c_2\, R[\tilde G]\, \L_{4}^{2n-2} + ... \, \)\,
\label{S-oneloop-gravity}
\ee
where $c_m$ are model-dependent constants, omitting 
dilaton-like terms. 
This allows already to draw some qualitative conclusions,
focusing on $2n=4$ --dimensional NC branes. 
More detailed computations should be 
performed elsewhere.

Note that we associated different scales $\L_1$ resp. $\L_4$ 
to the different terms in \eq{S-oneloop-gravity}, which 
arise as follows. It is well-known that the 
coefficient of the leading ``would-be cosmological constant'' 
term $\int d^4 x\,\sqrt{|\tilde G_{\mu\nu}|}$ 
is determined by the scale $\L_1$ for 
$N=1$ SUSY breaking. In contrast, the coefficient of the 
induced Einstein-Hilbert term in emergent NC gravity
is determined by the scale $\L_{4}$ 
where $N=4$ SUSY is broken. This reflects the well-known
fact that UV/IR mixing in NC gauge theory persists even in 
SUSY gauge theory \cite{Matusis:2000jf,Khoze:2000sy}, 
except in the $N=4$ case. 
Since  $\Gamma_{1-loop}[\tilde G]$ is nothing but a
re-interpretation of the UV/IR mixing terms in
NC gauge theory \cite{Steinacker:2007dq,Grosse:2008xr}, it follows 
that the induced term $R[\tilde G]$ has a cutoff given by the 
scale of $N=4$ SUSY breaking; this is discussed in \cite{Klammer:2008df} 
from the point of view of gravity.
Because there is no bare gravity action, it follows
that the effective Newton constant resp. the Planck scale
in emergent gravity is given by 
\be
l_{\rm Pl}^2 = \frac 1{G} \,\sim\, \L_{4}^2 \, .
\label{Newton-cutoff}
\ee  
This also suggests what happens in models without a finite
effective cutoff $\L_4$: Then $G \to 0$, hence the  the induced gravitational
action becomes a constraint and there is no more back-reaction 
of matter to the geometry. However there might still be interesting
scaling limits.

\subsection{Effective action and the (ir)relevance of the cosmological
constant term}

Consider again the case of 4-dimensional NC branes embedded in $\R^D$.
The full semi-classical effective action of the matrix model at one loop 
 is given by
\be
S_{eff} = S_{YM} + S_{1-loop}
\label{S-eff-oneloop}
\ee
where $S_{1-loop} = \Gamma_{1-loop}[\tilde G] + ...$, and
\bea
S_{YM} &=& \frac 1{(2\pi)^2 g^2}\, \int d^4 x\,
\sqrt{|\tilde G_{\mu\nu}|}\, \(e^{\sigma}\,\tilde G^{\mu\mu'} \tilde G^{\nu\nu'}
tr F_{\mu\nu}\, F_{\mu'\nu'}
+  \tilde G^{\mu\mu'} tr \partial_\mu\phi^i\partial_\nu\phi^j\d_{ij} \)  \nn\\
&&  + \frac 1{(2\pi)^2 g^2}\, \int d^4 x\,
 (4\rho \eta  + 2  \eta(x) \, tr F\wedge F\,)
\eea
including nonabelian gauge fields 
and the nonabelian components $\phi^i = \phi^i_\a \la^\a$  of the scalar fields.
Fermionic terms are omitted. We introduced 
an explicit coupling constant $g$, which does not enter the induced 
gravitational action 
since it can be absorbed in the fields. The 
bare YM coupling constant is given by 
$g_{YM}^2 = g^2 e^{-\sigma}$ \eq{YM-coupling-effective}, which
receives the standard quantum corrections, and
might play a role similar to a GUT coupling.
The one-loop induced gravitational term 
$\Gamma_{1-loop}[\tilde G]$  for 4-dimensional NC
branes is \eq{S-oneloop-gravity} 
\be
\Gamma_{1-loop}[\tilde G] 
\sim \int d^4 x\, \sqrt{|\tilde G_{\mu\nu}|}\; (\L_1^4
 + R[\tilde G]\L_4^2) 
\label{S-oneloop-gravity-4D}
\ee
where $|\tilde G_{\mu\nu}| = |g_{\mu\nu}|$ using \eq{G-g-4D}.
The first term is therefore simply the invariant volume of the embedding
metric. 

Consider the geometric equations of motion. It is well-known that 
\be
\d  \int d^4 x\, \sqrt{|\tilde G|}\;  R[\tilde G] 
 = \int d^4 x\, \sqrt{|\tilde G|}\;
\Big(\frac 12 R[\tilde G] \tilde G^{\mu\nu}  - R^{\mu\nu}[\tilde G]\Big)
\d \tilde G_{\mu\nu}
\label{variation-EH}
\ee
while the variation of the ``would-be cosmological term'' is  
\be
\d  \int d^4 x\, \sqrt{|\tilde G|} =  
\frac 12 \int d^4 x\, \sqrt{|\tilde G|}\, \tilde G^{\mu\nu}\,\d \tilde G_{\mu\nu}
= \frac 12 \int d^4 x\, \sqrt{g}\, g^{\mu\nu}\,\d g_{\mu\nu}\, ,
\label{variation-CC}
\ee
using \eq{G-g-4D}
in the case of 4-dimensional NC branes. This vanishes
identically in the case $D=4$ due to \eq{metric-unimod}.
The variation of the 
bare gravitational action $\d \int d^4 x\, 4\rho\eta$
was worked out in \eq{variation-bare-geometry}.

In General Relativity, \eq{variation-EH} and \eq{variation-CC}
imply the Einstein equations
for vacuum.
The essential difference here is that the metric $\tilde G_{\mu\nu}$
is constrained, and the fluctuations do not span the space of 
symmetric $4\times 4$ matrices. Therefore we do {\em not} simply
obtain the Einstein equations; this is seen most strikingly for
the cosmological constant term as discussed below. However, 
note that Ricci-flat spaces which can be realized in terms 
of our $\tilde G_{\mu\nu}$ certainly
satisfy $\d  \int d^4 x\, \sqrt{|\tilde G|}\;  R[\tilde G]  = 0$. This 
supports the physical viability of this framework.
The correct e.o.m. which follow from the effective action
are complicated here by the presence of possible dilaton-like terms
at one loop, and will be derived elsewhere. 
They will in particular modify \eq{eom-geom-covar-extra},
which takes into account the bare action only.

Let us briefly discuss the geometrical degrees of freedom.
We can decompose the variations $\d X^a$ of the basic matrices into 
tangential and normal fluctuations w.r.t. the background brane $\cM$. 
Using an orthogonal transformation if necessary, we can assume that 
$\d X^\mu \in T\cM$ are tangential, 
and $\d\phi^i \in T\cM^\perp$ are normal to
the brane at some given point. Consider first the tangential
variations $\d X^\mu = \cA^\mu$. 
They lead to  variations of the 
Poisson tensor $\theta^{\mu\nu}$ on the brane (which can be 
interpreted as diffeomorphism), but they do {\em not}
change the embedding metric, which is fixed in the matrix model
(recall that e.g.
$g_{\mu\nu} \equiv \eta_{\mu\nu}$ in the simplest case $D=4$):
$\d_\cA\, g_{\mu\nu} =0$.
Therefore these tangential fluctuations 
imply nontrivial\footnote{except for gauge transformations resp. 
symplectomorphisms $\cA^\mu = [f,X^\mu]$.} physical fluctuations 
of the effective metric $\d_\cA\, G_{\mu\nu} \sim (G \theta F + F
\theta G)_{\mu\nu} $ \eq{h-abdown}
corresponding to the 2 on-shell
graviton helicities plus one off-shell deformation;
cf. the discussion in Section \ref{sec:linearized}.
In particular, the 
 term $\int d^4 x\, \sqrt{|\tilde G|} = \int d^4 x\, \sqrt{|g|}$ in 
\eq{S-oneloop-gravity} is
{\em independent} of these tangential degrees of freedom.
This provides (part of) 
a mechanism for avoiding the cosmological constant problem.

Now consider the normal fluctuations $\d\phi^i$ of the brane
embedding, which in general imply nontrivial physical fluctuations 
of the effective metric. The corresponding  variation of the 
``would-be cosmological term'' is
\bea
\d \int d^4 x\,\sqrt{|g_{\mu\nu}|} 
&=&\int d^4 x\, \sqrt{|g|}\, g^{\mu\nu} \d g_{\mu\nu} 
=  2\int d^4 x\, \sqrt{|g|}\,g^{\mu\nu} 
\partial_\mu\phi^i \partial_\nu\d\phi^j \d_{ij}    \nn\\
&=& - 2\int d^4 x\,\sqrt{|g|} \, \d\phi^i  \Delta_g \phi^j \d_{ij}
\label{variation-det-g-2}
\eea
using partial integration, where $\Delta_g$ 
is the covariant Laplacian corresponding
to the metric $g_{\mu\nu}$. 
This vanishes if the $\phi^i$ satisfy the constraint
\be
\Delta_g \phi^i =0 \, ,
\label{laplace-g-phi}
\ee
which is similar to \eq{eom-phi} except that the 
metric is now $g_{\mu\nu}$. 
Flat embeddings do satisfy this condition. 
Therefore flat space is a solution
at the quantum level, even in the presence of this  ``would-be cosmological 
constant'' term; the same applies to any surfaces embedded in $\R^D$
which satisfies \eq{laplace-g-phi}.
This can also be seen from the fact that there is no tadpole
contribution at one loop in NC gauge theory
\cite{Grosse:2008xr}.
This is in stark contrast to General Relativity,
where the term
$\int d^4 x\sqrt{\tilde G}\, \L^4$
corresponds to a huge cosmological constant, 
requiring unreasonable fine-tuning.
Together with the observations of the previous paragraph,
we obtain strong evidence here that the
cosmological constant problem is resolved or at least much milder in
the present context. This is a robust mechanism,
rooted in the fact that the metric is not a fundamental degree of freedom
but emerges from the matrix model.

To obtain a more complete understanding of the
cosmological constant issue in emergent NC gravity,
a more complete analysis 
is required, as for related claims in the literature 
\cite{Yang:2007as}. At present, the only known solution 
for the full effective action \eq{S-eff-oneloop}
is flat Moyal-Weyl space.
Nevertheless, the fact that this is a solution 
without fine-tuning a cosmological constant is very remarkable.
Moreover,
since the Einstein-Hilbert term contains two explicit derivatives,
the bare action together with the
``would-be cosmological constant'' term will govern the extreme IR
(cosmological) scale which should indeed be flat, 
while the induced E-H action will 
determine the gravitational fields due to localized (point) masses. 
This would be a very satisfactory picture.

\section{Linearized metric and 
gravitational waves}
\label{sec:linearized}

\paragraph{Moyal-Weyl case.}

A particular solution of the e.o.m.
\eq{eom-Y} is given by the 4D Moyal-Weyl quantum plane.
Its generators $\bar X^{\mu}$  satisfy
\be
[\bar X^\mu,\bar X^\nu] = i \bar\theta^{\mu\nu} \one\, ,
\label{Moyal-Weyl}
\ee
where $\bar\theta^{\mu\nu}$
is a constant antisymmetric tensor. 
The effective geometry \eq{effective-metric}
 for the Moyal-Weyl plane is indeed
flat, given by
\bea
\bar g^{\mu\nu} &=& \bar\theta^{\mu\mu'}\,\bar\theta^{\nu\nu'} 
\eta_{\mu'\nu'}, \nn\\
\tilde g^{\mu\nu} &=& \bar \rho\, \bar g^{\mu\nu}   \nn\\
\bar\rho &=& |\bar g_{\mu\nu}|^{1/4} 
= |\bar\theta_{\mu\nu}^{-1}|^{1/2} \equiv \L_{NC}^4 .
\label{effective-metric-bar}
\eea
In this section,
lower-case $\bar g^{\mu\nu}$ resp. $\tilde g^{\mu\nu}$
 will denote the flat effective Moyal-Weyl metric
rather than the embedding metric, and
we will rise and lower indices using  $\bar g_{\mu\nu}$.
First, we can choose coordinates where 
$\tilde g_{\mu\nu} = (-1,1,1,1)$, so that $x^0 = c t$ 
corresponds to the time. 
One can use the 
remaining $SO(3,1)$ (resp. $SO(4)$ in the Euclidean case)
to bring $\bar \theta^{\mu\nu}$ into canonical form
\be
\bar\theta^{\mu\nu} = \theta\left(\begin{array}{cccc} 0 & 0 & 0 & -1 \\
                                0 & 0 & \a & 0 \\
                                0 & -\a & 0 & 0  \\
                                1 & 0 & 0 & 0 \end{array}\right)\, .
\label{theta-standard}
\ee 
The NC scale is then
\be
\bar\rho = \theta^{-2}\a^{-1} = \L_{NC}^4 ,
\ee
and the original flat background metric is
\be
\eta_{\mu\nu} = \bar \theta^{-1}_{\mu\mu'} \bar \theta^{-1}_{\nu\nu'} 
\bar g^{\mu'\nu'} 
= \bar \rho^{-1}\,\theta^{-2}\diag(1,\a^{-2},\a^{-2},-1) 
= \a\, \diag(1,\a^{-2},\a^{-2},-1) \, .
\ee
The bare action for this flat Moyal-Weyl background is 
\be
S_{YM} = Tr \,\bar\theta^{\mu\nu} \bar\theta^{\mu'\nu'} 
\eta_{\mu\mu'}\eta_{\nu\nu'} 
= Tr \,\eta_{\mu\nu}\bar g^{\mu\nu} 
= \int d^4 x \,\eta_{\mu\nu}  \tilde g^{\mu\nu} 
= 2\int d^4 x\, \a\, (\a^{-2}-1) 
\ee
which vanishes in the case $\a=1$ where
$\bar\theta^{\mu\nu}$ admits an enhanced $SO(2,1) \times U(1)$ symmetry.
In the Euclidean case, the action is positive definite. 
From now on, we assume that 
$$\a=1$$ 
for simplicity. It is also worth pointing out that  
we are not in the case of ``space-like'' 
noncommutativity, since $\bar \theta^{\mu\nu}$ is 
non-degenerate.
However, the problems of unitarity etc. discussed e.g. in 
\cite{Gomis:2000zz} 
are expected to be benign 
in the present context due to the assumed $N=4$
supersymmetry at the Planck scale.

\paragraph{Deformations of the flat Moyal-Weyl plane.}

Consider small deformations of the flat Moyal-Weyl plane,
\be
X^\mu = \bar X^{\mu} -\bar\theta^{\mu\nu} A_\nu(x) \, ,
\label{cov-coord-1}
\ee
where $A_\nu$ are hermitian and can be interpreted as $U(1)$ gauge fields
on $\R^{4}_\theta$ with field strength 
$\bar F_{\mu\nu} = \partial_\mu A_\nu - \partial_\mu A_\nu + i[A_\mu,A_\nu]$.
The linearized metric $G^{\mu\nu}$  is 
\bea
G^{\mu\nu} &=&
 (\bar \theta^{\mu \mu'} - \bar \theta^{\mu \eta}\bar \theta^{\mu' \rho}\, 
 \bar F_{\eta \rho}) 
 (\bar \theta^{\nu \nu'} - \bar \theta^{\nu \sigma}\bar\theta^{\nu'\kappa}\,
 \bar F_{\sigma \kappa})
(\eta_{\mu'\nu'} + \d g_{\mu'\nu'}) \, \nn\\
&=& \bar g^{\mu\nu} -h^{\mu\nu} 
\label{G-general}
\eea
where
\be
h^{\mu\nu} = - \bar g^{\mu \mu'}\, \bar F_{\mu' \eta}\bar \theta^{\eta\nu}
- \bar \theta^{\mu \mu'}\, \bar F_{\mu' \eta}\,\bar g^{\nu \eta} 
- \bar \theta^{\mu \mu'}\bar \theta^{\nu \eta}\d g_{\mu' \eta}
\quad + O(A^2)\, .
\label{graviton-general}
\ee
where $\d g_{\mu\nu} = \partial_\mu \phi^i \partial_\nu \phi^j\,\d_{ij}$.
Correspondingly, the inverse metric is
\be
G_{\mu\nu} = \bar g_{\mu\nu} + h_{\mu\nu} + \dots\,,
\ee
with
\be
h_{\mu\nu} \equiv  \bar g_{\mu \mu'} \bar g_{\nu \nu'} h^{\mu'\nu'}
 =  - \bar g_{\nu \nu'} \bar\theta^{\nu'\rho} \bar F_{\rho\mu} 
- \bar g_{\mu \mu'} \bar\theta^{\mu'\rho} \bar F_{\rho\nu} \, 
-  \bar g_{\mu \mu'} \bar g_{\nu \nu'} \bar \theta^{\mu'\rho'}
\bar \theta^{\nu'\eta'}\d g_{\rho'\eta'} .
\label{h-abdown}
\ee
This gives
\be
h = h_{\mu\nu} \obar g^{\mu\nu} = 2 \obar \theta^{\mu\nu}\, F_{\mu\nu}\,
 - \eta^{\mu\nu} \d g_{\mu\nu}\, .
\label{h-wave}
\ee
We focus on the case of flat embeddings $\d g_{\mu\nu} =0$. Then
$h^{\mu\nu} = - \bar g^{\mu \mu'}\, \bar F_{\mu'\rho}\bar \theta^{\rho\nu}
- \bar \theta^{\mu \mu'}\,\bar F_{\mu'\rho}\,\bar g^{\nu \rho}\,\,+ O(A^2)$
gives the linearized fluctuation resp. graviton in terms of the 
$U(1)$ degrees of freedom.
The linearized Ricci tensor for the unimodular
metric $\tilde G_{\mu\nu}$ resp. the traceless graviton
$\tilde h_{\mu\nu} = h_{\mu\nu} -\frac 14 \obar g_{\mu\nu} h$ is given by
\be
R_{\mu\nu}[\tilde G] =
  \frac 12\({\obar \theta_\mu}^{\eta}\,
\partial^\rho \partial_{\eta} F_{\rho\nu}
+  {\obar\theta_\nu}^{\eta}\, \partial^\rho\partial_{\eta}  F_{\rho\mu}
+\frac 12 \obar g_{\mu\nu}\partial^\rho \partial_\rho 
F_{\eta\sigma}\bar\theta^{\eta\sigma}\)\, 
\label{ricci-unimodular}
\ee
in agreement with results of \cite{Rivelles:2002ez}, and
\be
R[\tilde G] 
= \frac 12 \partial^\rho \partial_\rho\,  
\obar \theta^{\eta\sigma} F_{\eta\sigma} \, .
\ee
Now consider the equations of motion for the bare action 
\eq{eom-Y-2},
which in the present context amount to 
$\partial^\mu F_{\mu\nu}=0 = \partial^\rho\partial_{\rho}  F_{\mu\nu}$
up to possibly corrections of order $\theta$,
i.e. the vacuum Maxwell equations for the flat metric $\obar g_{\mu\nu}$.
As pointed out in \cite{Rivelles:2002ez}, this implies that
the vacuum geometries are Ricci-flat to leading nontrivial order,
\be
R_{ab}[\tilde G] = 0 + O(\theta^2), 
\label{ricci-flat}
\ee
while the general curvature tensor $R_{\mu\nu\rho\eta}$ is first order in
$\theta$ and does not vanish\footnote{while this is true 
generically, there may be 
particular momenta $k^\mu$ determined by $\theta^{\mu\nu}$ for which
the corresponding ``graviton'' 
is pure gauge and hence $R_{\mu\nu\rho\sigma}$ vanishes. 
This should be studied in more 
detail elsewhere.}. This
shows that the effective metric does contain the 2 physical 
degrees of freedom (helicities) of
gravitational waves. It is quite remarkable that \eq{ricci-flat} is obtained 
from the bare action,
without invoking the mechanism of induced gravity 
in section \ref{sec:induced-grav}. Note that the
cosmological constant vanishes to this order.
A generalization to non-trivially embedded branes remains to be 
elaborated.

\subsection{Newtonian Limit and relativistic corrections}
\label{sec:newton-limit}

The Newtonian limit of General Relativity corresponds to static metric
perturbations of the form 
\be
ds^2 = -c^2 dt^2\Big(1+\frac {2U}{c^2}\Big) 
+  d \vec x^2 \Big(1+O(\frac 1{c^2})\Big)
\label{newton-metric}
\ee
where $\Delta_{(3)} U(x) = 4\pi G m(x)$ and $m(x)$ denotes the mass density. 
Including the leading relativistic corrections, this takes the form
\be
ds^2 = -c^2 dt^2\Big(1+\frac {2U}{c^2}\Big) 
+  d \vec x^2 \Big(1-\frac {2U}{c^2}\Big)\, .
\label{newton-metric-corr}
\ee
It follows from the results of the previous section that 
\eq{newton-metric-corr}
is reproduced correctly in the case without matter, 
where the vacuum equations of motion 
amount to $\partial^c F_{cb} =0$ resp. $R_{ab}=0$.
In the presence of matter, it was essentially shown in 
\cite{Steinacker:2007dq} that 
one can indeed obtain metrics of the form 
\eq{newton-metric} for arbitrary static $m(x)$.
We will re-analyze this issue  in more detail here. 
It will turn out that even though
one can find a metric $\tilde G_{\mu\nu}$ corresponding to 
\eq{newton-metric} in the case $D=4$ without nontrivial
embeddings, the relativistic corrections of General Relativity 
in \eq{newton-metric-corr}
are not correctly reproduced in the presence of matter. In particular,
it appears that the Schwarzschild solution is not correctly reproduced in this minimal
framework. This is not a real problem for emergent NC gravity, since we concluded on 
different grounds above that the  model with $D=10$ and
$N=4$ SUSY is required for a consistent model at the quantum level. 
Therefore realistic solutions for point masses should be realized by
nontrivially embedded branes.

In order to reproduce \eq{newton-metric} with $\tilde G_{\mu\nu}$, 
we have to 
find $U(1)$ gauge fields $A_\mu$ on the Moyal-Weyl quantum plane 
with the desired $h_{\mu\nu}$, which is  \eq{h-abdown}
\be
h_{\mu\nu} = - \obar g_{\nu \nu'}\obar \theta^{\nu'\eta}\, F_{\eta\mu}
- \obar g_{\mu \mu'}\obar \theta^{\mu'\eta}\, F_{\eta\nu} \, .
\label{flat-grav-wave-2}
\ee
Choose coordinates where 
$\tilde g_{\mu\nu} = (-1,1,1,1)$ as discussed above, so that $x^0 = c t$ 
corresponds to the time, and $\bar\theta^{\mu\nu}$ 
has the  form \eq{theta-standard}.
Then 
\be
F_{\mu\nu} = \left(\begin{array}{cccc} 0 & E_1 & E_2 & E_3  \\
                                -E_1 & 0 & B_3 & -B_2 \\
                                -E_2 & -B_3 & 0 & B_1  \\
                                -E_3 & B_2 & -B_1 & 0
                              \end{array}\right) \, 
\ee
gives
\be
h_{\mu\nu} = \bar\rho\theta
    \left(\begin{array}{cccc} -2 E_3 & B_2- E_2 & -B_1+ E_1 & 0 \\
                              B_2-  E_2 & -2  B_3 & 0 &  B_1+E_1 \\
                            -B_1+  E_1 & 0 & -2  B_3 &  B_2+E_2  \\
                            0 &   B_1+E_1 &  B_2+E_2 & 2E_3
                          \end{array}\right) 
\ee
which is the most general metric fluctuation available.
Let us denote its trace with
\be
h(x) = \bar g^{\mu\nu} h_{\mu\nu}(x) 
=  4 \theta (E_3 - B_3)  \, .
\label{h}
\ee
The physical graviton is the traceless version,
\be
\tilde h_{\mu\nu} = h_{\mu\nu} - \frac 14 \bar g_{\mu\nu} h 
= \bar\rho\theta
    \left(\begin{array}{cccc} -(B_3+ E_3)  & B_2- E_2 & -B_1+ E_1 & 0 \\
                              B_2-  E_2 & - (B_3+E_3) & 0 &  B_1+E_1 \\
                            -B_1+  E_1 & 0 & -(B_3+E_3) &  B_2+E_2  \\
                            0 &   B_1+E_1 &  B_2+E_2 & B_3 + E_3 \end{array}\right).
\label{h-traceless-explicit}
\ee
As shown above, $R_{\mu\nu}[\tilde G] =0$ holds if 
$\vec E$ and $\vec B$ 
satisfy the Maxwell equations without source. We would like this 
to be the case where the mass density $m(x)$ vanishes.

\subsubsection{Static point charges}
\label{sec:pointcharge}

To gain some intuition, we first consider a static
point charges with electric and magnetic charge at the origin, and 
determine the corresponding effective geometry.
Thus consider electromagnetic fields given by
\be
E_i =  q_E \frac 1{r^3}\, x_i, \qquad
B_i =  q_M \frac 1{r^3}\, x_i .
\ee
Then the metric fluctuation \eq{h-traceless-explicit} is 
\be
\tilde h_{\mu\nu} = \frac 1{r^3}\,\bar\rho\theta
    \left(\begin{array}{cccc} -(q_E+q_M) x_3  & (q_M-q_E) x_2 & -(q_M-q_E) x_1 & 0 \\
                           (q_M-q_E) x_2  & -(q_E+q_M) x_3  & 0 & (q_M+q_E) x_1 \\
                          -(q_M-q_E) x_1  & 0 & -(q_E+q_M) x_3  & (q_M+q_E) x_2  \\
                          0 & (q_M+q_E) x_1  & (q_M+q_E) x_2  &
                          (q_E+q_M) x_3 \end{array}\right)\, ,
\label{h-traceless-explicit-point}
\ee
which in the ``extremal'' case $q_M = q_E =:q$ is
\be
\tilde h_{\mu\nu} = \frac{2q\bar\rho\theta}{r^3}\,
    \left(\begin{array}{cccc} - x_3  & 0 & 0 & 0 \\
                           0  & -  x_3  & 0 &  x_1 \\
                            0  & 0 & - x_3  &  x_2  \\
                           0 &  x_1  &  x_2  &  x_3
                         \end{array}\right)\, .
\label{h-traceless-explicit-point-extr}
\ee
This can be brought into diagonal form using the diffeomorphism
$\xi^\mu =2q\bar\rho\theta (0,0,0, \frac 1r)$, 
which gives the linearized metric
\bea
\tilde h_{\mu\nu}' &=& \tilde  h_{\mu\nu} 
+ \partial_\mu \xi_b + \partial_\nu \xi_a \nn\\
&=&\frac{2q\bar\rho\theta}{r^3}\,
    \left(\begin{array}{cccc} - x_3  & 0 & 0 & 0 \\
                           0  & -  x_3  & 0 &  0 \\
                            0  & 0 & - x_3  &  0  \\
                           0 & 0  &  0 &  -x_3 \end{array}\right)
\label{pointcharge-h-explicit}
\eea
This has indeed the form of \eq{newton-metric-corr} of a 
Ricci-flat metric for $\vec x \neq 0$, 
with Newtonian potential 
$U =  \frac{q\bar\rho\theta}{r^3}\,x_3 \sim \partial_3 \frac 1r$
which is harmonic away from the origin. However, the corresponding
``mass'' distribution 
\be
m(x) \sim \Delta U \sim  \partial_3 \delta^{(3)}(\vec x)
\ee
is not positive. This is not what we want;
it corresponds to an unphysical gravitational dipole
rather than a point mass. Note however that
{\em there is no charged field in the model} for this 
$U(1)$, hence this is only a toy configuration which is
not expected to play any physical role. Moreover,
it is not expected to be a solution of the e.o.m.
at the quantum level.

This  result is easy to understand:
Since the electromagnetic field of a localized charge 
distribution decays as $\frac 1{r^2}$, the corresponding 
gravitational field also decays like  $\frac 1{r^2}$ at the linearized
level.  
The correct $U(r) \sim \frac 1r$ 
gravitational potential for a point mass
can be recovered either at the cost of violating relativistic 
corrections as elaborated next, or -- presumably -- by a
nontrivial deformation of the brane embedding 
due to the point mass, governed by the induced
gravitational action.
The trace-$U(1)$ modes under consideration here are to be interpreted
as gravitational waves.

\subsubsection{General mass distributions}

Now consider more generally the static case 
\be
\tilde h_{0i} = 0, \qquad \partial_0 \tilde h_{\mu\nu} =0\, ,
\ee
which amounts to 
\be
B_2 =  E_2, \quad B_1 =  E_1 \, .
\label{static}
\ee
The metric fluctuation \eq{h-traceless-explicit} is then 
\be
\tilde h_{\mu\nu} = \bar\rho\theta
    \left(\begin{array}{cccc} -(B_3+E_3) & 0 & 0 & 0 \\
                              0 & - (B_3+E_3) & 0 & 2 E_1 \\
                              0 & 0 & - (B_3+E_3) & 2 E_2  \\
                         0 & 2 E_1 & 2E_2 & B_3+E_3 \end{array}\right)\, .
\ee
To determine the covariant coordinates $x^\mu = \bar x^\mu -
\bar\theta^{\mu\nu} A_\nu$ \eq{cov-coord-1},
we have to fix a gauge.
A natural gauge choice in the present context is the ``static gauge'' 
$\partial_0 A_\mu=0$, so that\footnote{alternatively one can 
also impose e.g. $A_0=0$, but then the $A_i$ 
become $x^0$-dependent \cite{Steinacker:2007dq}.}
\be
\vec E = -\partial_i A_0(\vec x), \qquad
\vec B = \vec\nabla\times\vec A(\vec x) \, .
\label{static-gauge}
\ee
The metric can be brought into diagonal form
using the diffeomorphism
${x^\mu}' = x^\mu +  \xi^\mu(x)$ with $\xi^\mu(x) = 2\bar\rho\theta(0,0,0,A_0(x))$,
which gives
\bea
\tilde h_{\mu\nu}' &=& \tilde h_{\mu\nu} + \partial_\mu \xi_\nu + \partial_\nu \xi_\mu
 = \bar\rho\theta
    \left(\begin{array}{cccc} -(B_3+E_3) & 0 & 0 & 0 \\
                              0 & -(B_3+E_3) & 0 & 0 \\
                              0 & 0 & -(B_3+E_3) & 0  \\
                             0 & 0 & 0 & B_3-3 E_3  \end{array}\right)\nn\\
&=& -2 U(x) \one 
-  \frac 12\bar\rho\,
    \left(\begin{array}{cccc} 0  & 0 & 0 & 0 \\
                              0 & 0 & 0 & 0 \\
                              0 & 0 & 0 & 0  \\
                             0 & 0 & 0 & h(x) \end{array}\right) 
\label{h-tilde-general-diag}
\eea
where we used \eq{h} to write 
$E_3 =   B_3 + \frac 1{4\theta}\, h$, hence
\be
\vec B =  \vec E + 
\(\begin{array}{c}0\\0\\ - \frac 1{4\theta} h(x)
 \end{array}\) .
\ee
For $h(x)=0$, this has precisely the form of the metric \eq{newton-metric-corr}
with Newtonian potential  
\be
2U  =  \bar\rho\theta\, (B_3 + E_3)
 =  -\bar\rho (2\theta\, \partial_3 A_0 + \frac 1{4}\, h) \, 
\label{newton-E}
\ee
including leading relativistic corrections,
while for $h\neq 0$ it agrees with the Newtonian limit  \eq{newton-metric}
but the last term in \eq{h-tilde-general-diag}
violates the relativistic corrections.

To determine $A_\mu$ explicitly for given $U(x)$, we act 
with $\partial_3$ on \eq{newton-E} and combine the result with
the Bianci identity for $\vec B$ 
\be
0 = \div \vec B =  \div\vec E 
 - \frac 1{4\theta} \partial_3 h(x)
= -\Delta A_0(x) - \frac 1{4\theta} \partial_3 h(x)  \, .
\label{bianci-1}
\ee
This gives
\be
\partial_3 \frac{2U(x)}{\bar\rho\theta}
 = -2 \partial_3^2 A_0  +\Delta A_0(x)\, 
 = (-\partial_3^2 + \partial_1^2+ \partial_2^2) A_0 ,
\label{A-0-equation}
\ee
which can be solved for $A_0$ as
\be
A_0(x) = \frac{2}{\bar\rho\theta} 
\int d^3 y\, G(x-y) \frac{\del}{\del y^3}U(y)\, .
\label{A0-solution-propagator}
\ee
Here $G(x-y)$ is a 3-dimensional propagator,
$(-\partial_3^2 + \partial_1^2+ \partial_2^2)G(x-y) =
\delta^{(3)}(x-y)$. The (static) Bianci identity for $\vec E$ 
\be
0 =  \rot \vec E = \rot\vec B 
 + \frac 1{4\theta} (\partial_2 h(x),-\partial_1 h(x),0) 
\label{bianci-2}
\ee
then determines the conserved current 
\be
\vec J \equiv \rot \vec B 
= -\frac 1{4\theta} (\partial_2 h,-\partial_1 h,0),
\qquad \div \vec J  = 0 \, ,
\ee
so that $\vec B = \vec\nabla\times\vec A$ 
can be solved for $\vec A$.
Therefore for an arbitrary given potential $U(\vec x)$,
we can indeed find $A_\mu$ corresponding to a static 
 metric fluctuation $h_{\mu\nu}$
which reproduces the Newtonian potential $U(\vec x)$.
There is some freedom in the solution of $A_0$ 
\eq{A-0-equation}, and $h(x)$ is (almost) determined by \eq{bianci-1}.
Note that 
even though a preferred direction $x^3$ is singled out through
$\theta^{\mu\nu}$ and $x^0$, this merely amounts to 
preferred coordinates $x^\mu$ for the
desired geometry.

Let us consider the vacuum case $\Delta U(x) =0$ in more detail. 
By integrating \eq{newton-E},
we can obtain a Ricci-flat solution with $A_i =0,\, h(x) =0$,
\be
A_0(x) =  \frac 1{\theta}\,\int_{0}^{x^3} ds\, U(x^1,x^2,s) 
 + H(x^1,x^2)
\ee
which solves $\Delta A_0 =0$ provided
 $(\partial_1^2 + \partial_2^2) H 
= -\frac 1{\theta}\,\frac{\del}{\del x^3}U(x)|_{x^3 =0}$.

Now consider the case with non-vanishing mass distribution 
$\Delta U(x) = 4\pi G m(x) \neq 0$
in a region of space near the origin. Then 
the presence of a (hyperbolic!) propagator
in \eq{A0-solution-propagator} implies that $A_0$ will not be harmonic
even in regions where the mass density $m(x)$ vanishes. This in turn
implies e.g. through \eq{bianci-1} that $h(x) \neq 0$, and the 
leading relativistic
corrections to Newtonian gravity are not correctly reproduced.
In other words, while it is possible to reproduce 
e.g. the Newtonian potential
$U(r) \sim \frac 1r$ for a point mass, it implies in the
electromagnetic picture a nontrivial charge density which is not 
localized at the origin, leading to a violation of 
Ricci flatness. This is in accord with the results of Section 
\ref{sec:pointcharge}. 

We conclude that the consideration 
of nontrivially embedded branes in matrix models with extra dimensions
is required in order to obtain a gravity theory which reproduces
the leading relativistic corrections of General Relativity
in the presence of masses. This is in accord with
the results of Section \ref{sec:induced-grav} 
that $N=4$ SUSY is required at the 
quantum level, leading to the $D=10$ IKKT model and hence  
to embedded branes.
Since the embedding  degrees of freedom can be viewed as 
scalar fields, their quantization is straightforward, 
and expected to be well-behaved.

\section{Solution with spherically symmetric Poisson structure}
\label{sec:spherical}

In this section, we discuss an exact but unphysical solution of the
tree-level e.o.m. \eq{eom-geom-covar-extra}, 
in order to illustrate nontrivial geometries and the covariant 
formulation.
The solution is unphysical, because the induced gravity action is 
not taken into account; this will be explored elsewhere.
We start from the covariant e.o.m. \eq{eom-covar-2} for 
the Poisson structure $\theta^{\mu\nu}$
\be
\tilde G^{\g \eta}(x)\, \tilde\nabla_\g \theta^{-1}_{\eta \nu} 
 = - \tilde G^{\g \eta}(x)\, \theta^{-1}_{\eta \nu}
\partial_\g \sigma + 
e^{-2\sigma}\,\tilde G_{\mu\nu} \theta^{\mu\g}\,\partial_\g\eta(x)
\label{maxwell-covar} 
\ee
and look for a solution 
$\theta^{-1}_{\mu\nu}$ which is static and spherically
symmetric. This Ansatz is actually {\em not} appropriate
in order to
look e.g. for a Schwarzschild-like solution; for that purpose
one should  presumably look for a deformation of the flat 
Moyal-Weyl solution, where $\theta^{\mu\nu}$ breaks
rotational invariance. Nevertheless, finding a nontrivial
exact solution of \eq{maxwell-covar}
is certainly instructive.

To illustrate the case of nontrivially
embedded branes, consider a 4-dimensional brane 
$\cM^4 \subset \R^5$ in the 
matrix model \eq{YM-action-extra}, with Cartesian coordinates 
$x^a =(x^\mu,\phi)$ given by 
the semi-classical limit of the matrices $X^a$. We also
use the radial variable
$r^2 = x_1^2+x_2^2+x_3^2$ and the Euclidean time $\tau = x_4$.
This leads to the following spherically symmetric closed 2-form
\be
\theta^{-1} = \omega^{(2)} + f(r,\tau) dr\wedge d\tau
\label{symplect-rot}
\ee 
where $\omega^{(2)} = \sin(\theta) d\theta\wedge d\phi$ can be interpreted
as field of a magnetic monopole on $S^2$, which is singular at the 
origin.
This will define a spherically symmetric metric $G_{\mu\nu}(x)$ if 
the  induced metric $g_{\mu\nu}(x)$
on $\cM^4$ is spherically symmetric. Hence 
we consider an embedding function
$\phi = \phi(r)$, so that\footnote{A seemingly more general $\phi(r,\tau)$ 
could be reduced to the above through a redefinition of $\tau$} 
\bea
g_{\mu\nu}(x) &=& \d_{\mu\nu} + \partial_\mu\phi \partial_\nu \phi \nn\\
d s^2_g &=&  r^2 (d\theta^2 + \sin^2(\theta)d\varphi^2) 
+ (1+ \phi'^2(r))dr^2 + d\tau^2 
\label{g-withphi}
\eea 
or
\be
g_{rr} = 1+ \phi'^2(r), \quad g_{\tau\tau} = 1,
\quad g_{\theta\theta} = r^2, 
\quad g_{\varphi\varphi} = r^2 \sin^2(\theta)\, .
\ee
Since $\theta^{-1}_{\varphi\theta} = \sin(\theta)$,
this gives 
\bea
d s^2_G &=& r^{-2} (d\theta^2 + \sin^2(\theta)d\varphi^2)
+ f(r,\tau)^{2}\((1+ \phi'^2(r))^{-1} d\tau^2 + d r^2\) \, .
\eea
The effective metric $\tilde G_{\mu\nu} = e^{\sigma}\, G_{\mu\nu}$ is 
\be
d s^2_{\tilde G}
 = \frac{\sqrt{1+\phi'^2(r)}}{f(r,\tau)}\, 
(d\theta^2 + \sin^2(\theta)d\varphi^2) 
 + r^2 f(r,\tau)\(\frac{1}{\sqrt{1+ \phi'^2(r)}}\, d\tau^2 
 + \sqrt{1+\phi'^2(r)}\, d r^2 \) \nn\\
\label{effective-G-spherical}
\ee
where 
\be
e^{-\sigma} = \frac{|\theta^{-1}_{ab}|^{1/2}}{|g_{ab}|^{1/2}}
 = \frac{f(r,\tau)}{r^2 \sqrt{1+\phi'^2(r)}} 
\ee
and
\bea
\eta &=& \frac 14\, G^{ab} g_{ab} 
= \frac 12 \Big(\frac{1+\phi'^2(r)}{f^2(r,\tau)} +  r^4\Big) \, .
\eea
We could now define 
\bea
f(r,\tau) &=& r^{-2}, \nn\\
\sqrt{1+ \phi'^2(r)} &=& (1-\frac{R_S}{R})^{-1},  \nn\\
 R^2 (1-\frac{R_S}{R}) &=& r^2\,
\eea
which reproduces the (Euclidean) Schwarzschild metric. 
However, we are not free to choose 
the embedding function $\phi(r)$, which must satisfy
an e.o.m. which for the bare matrix model
is given by $\Delta_{\tilde G} \phi =0$ 
\eq{eom-phi}, modified by 
the quantum effective action, cf. \eq{laplace-g-phi}.
Therefore the above metric is only an illustration how
nontrivial geometries may be realized.
We leave this for future work, and proceed to
give an illustrative solution only for the case 
of flat embedding $\phi=0$ resp. $D=4$.

\paragraph{Flat embedding $\phi =0$, or $D=4$.}

Now consider the purely 4D case with $\phi =0$. 
The effective metric \eq{effective-G-spherical} then becomes
\be
\tilde G_{ab} 
 = \frac{1}{f(r,\tau)}\, (d\theta^2 + \sin^2(\theta)d\varphi^2) 
 + r^2 f(r,\tau)\(d\tau^2 + d r^2 \) 
\label{metric-4D}
\ee
where 
\be
e^{-\sigma} =  \frac{f(r,\tau)}{r^2} , \qquad
\eta = \frac 12 \Big(\frac{1}{f^2(r,\tau)} +  r^4\Big) \,.
\label{etarho-4D}
\ee
Computing the Christoffel symbols for 
the  metric \eq{metric-4D} gives
\bea
\tilde\Gamma^r_{rr} &=& 
 r^{-1} + \frac 12f(r,\tau)^{-1}\partial_r f(r,\tau) 
= - \tilde\Gamma^r_{\tau \tau} 
= \tilde\Gamma^\tau_{r\tau} \nn\\
\tilde\Gamma^r_{\tau r} &=& 
  \frac 12 f(r,\tau)^{-1}\partial_\tau f(r,\tau)
= - \tilde\Gamma^\tau_{\tau\tau}
= - \tilde\Gamma^\tau_{rr} \nn\\
\tilde\Gamma^r &=& - \tilde G^{rr} f\partial_r  f^{-1},
\qquad  \tilde\Gamma^\tau = - \tilde G^{rr} f\partial_\tau  f^{-1} \,.
\eea
The covariant Maxwell equations \eq{maxwell-covar} 
for the $\tau$ component is then
\be
 \tilde G^{rr}\,\partial_r f - \tilde \Gamma^r f
+ \tilde G^{\tau\tau}\,\tilde \Gamma_{\tau\tau}^r f 
- \tilde G^{rr}\,\tilde \Gamma_{r\tau}^\tau f \nn\\
= e^{-2\sigma}\,\tilde G_{\tau\tau}\theta^{\tau r}\,
 \partial_r\eta(x)
 - \tilde G^{rr}\,\theta^{-1}_{r\tau} \partial_r \sigma
\ee
which gives
\be
f(r,\tau)^{-1}\partial_r f(r,\tau) = - 2 f^2 r^3 
\label{eq-r-1}
\ee
using \eq{etarho-4D}.
Similarly, the $r$ component of \eq{maxwell-covar} 
\be
-\tilde G^{\tau\tau}\,\partial_\tau f + \tilde \Gamma^\tau f
- \tilde G^{r r}\,\tilde \Gamma_{rr}^\tau f 
+ \tilde G^{\tau\tau}\,\tilde \Gamma_{\tau r}^r f 
= - e^{-2\sigma}\,\tilde G_{rr} f(r,\tau)^{-1}\,
 \partial_\tau\eta(x)
 + \tilde G^{\tau\tau}\,f(r,\tau) \partial_\tau \sigma 
\ee
gives
\be
 f(r,\tau)^{-1}\partial_\tau f(r,\tau) 
= - e^{-2\sigma}\,r^4 \, \partial_\tau\eta(x)
 + \partial_\tau \sigma  = 0
\label{eq-tau-2}
\ee
which  implies $f = f(r)$.
Together with \eq{eq-r-1} we obtain
\be
f(r) = \frac 1{\sqrt{r^4 + c}}\,.
\ee
To make contact with the standard notation, denote
\bea
R^2 &=& f^{-1} = \sqrt{r^4 + c}, \nn\\
r^2 f &=&  \frac{r^2}{R^2} = \sqrt{1 - \frac{c}{R^4}} \, .
\eea
Then $R^3 dR = r^3 d r$, hence 
\be
\frac{dr}{dR} = \frac{R^3}{r^3}
\ee
and we obtain
\bea
r^2 f dr^2 &=&  \frac{r^2}{R^2} \,(\frac{dr}{dR})^2 dR^2 
= \frac{1}{1 - \frac{c}{R^4}} dR^2
\eea
Therefore  the effective metric \eq{metric-4D} becomes
\be
d s^2_{\tilde G} 
 = R^2\, (d\theta^2 + \sin^2(\theta)d\varphi^2) 
 + \sqrt{1 - \frac{c}{R^4}}\, d\tau^2 
 +  \frac{1}{1 - \frac{c}{R^4}} dR^2
\label{metric-4D-solution}
\ee
which is flat for $c=0$.
This metric, in particular 
the radial dependence is quite strange. 
However this is not surprising,
because it is a solution of the ''bare'' equations of motion 
only, without taking into account the induced gravitational 
action. Therefore this serves merely as an illustration
how nontrivial solutions can arise. Since
the $\theta^{-1}_{\mu\nu}$ we used is far
from the Moyal-Weyl plane, the results of Section
\ref{sec:linearized} do not apply, and there is no contradiction
with the fact that \eq{metric-4D-solution} is not
Ricci-flat. Indeed, note that
\be
e^{-\sigma} = \frac 1{r^2 R^2} \sim \frac 1{R^4}
\ee
which is far from the Moyal-Weyl case where $e^{-\sigma} = const$.
In particular, even the solution $f = r^{-2}$ with flat
$\tilde G_{\mu\nu}$ is very different from the Moyal-Weyl plane.
This shows how the same geometry may be 
realized in different ways. These different realizations 
however will be distinguished once the induced gravitational
action is taken into account, which includes in particular
an action for the dilaton-like field $\sigma$ 
\cite{Klammer:2008df}.

\paragraph{Cartesian coordinates.}

To clarify the above solution, reconsider the 
spherically symmetric symplectic form \eq{symplect-rot} for $\phi=0$. 
The most general rotationally invariant
antisymmetric tensor in 3+1 (Euclidean) dimensions has the form
\be
\theta^{-1}_{0i} = x_i\, f(r,\tau), 
\qquad \theta^{-1}_{ij} = \varepsilon_{ijk} x_k \, g(r,\tau)
\ee
which is actually also invariant under $SO(3)$. 
The corresponding 2-form
\be
\theta^{-1} =  f(r,\tau) \, x_i d\tau dx_i  
+  g(r,\tau) \varepsilon_{ijk} x_k d x_i dx_j
\ee
is closed if and only if
\be
g(r) = r^{-3},
\ee
recovering \eq{symplect-rot}
\be
\theta^{-1} =  f(r,\tau) \, x_i d\tau dx_i  
+  r^{-3} \varepsilon_{ijk} x_k d x_i dx_j \,
\ee
for an aritrary function $f(r,\tau)$.
The corresponding effective metric is 
\bea 
G_{00} &=& \theta^{-1}_{0i} \delta^{ij} \theta^{-1}_{0j} 
 = r^2 f^2 \nn\\
G_{ii} &=& \theta^{-1}_{ik} \delta^{kl} \theta^{-1}_{il} 
 + \theta^{-1}_{i0} \delta^{00} \theta^{-1}_{i0} 
= r^{-6} \delta_{ii} r^2 + x_i x_i (f^2 -  r^{-6})\, .
\eea
For $f = r^{-3}$, we obtain 
\be
G_{\mu\nu} = 
\frac 1{r^4}\,\(\begin{array}{ccc}
 1 & \vline & 0 \\
\hline  & \vline &  \\
 0 & \vline & \d_{ii} \end{array}\)\, ,
\ee
so that $\tilde G_{\mu\nu}$ reproduces the flat solution 
found above \eq{metric-4D-solution} for $c=0$.
Note again that the corresponding $\theta^{-1}_{\mu\nu}$
is very different from the Moyal-Weyl case
where $\theta^{-1}_{\mu\nu} = \rm {const}$.

We conclude that in order 
to obtain realistic metrics such as the Schwarzschild-metric,
different nontrivial embeddings must be used, solving the 
equations of motion derived from the 
combined bare action plus induced gravitational 
action.

\section{Symmetries and conservation laws}

The basic matrix model \eq{YM-action-extra} is invariant under 
the D-dimensional Poincar\'e group, consisting of translations
\be
X^a \to X^a + c^a, \qquad c^a \in \R
\label{translations}
\ee
and rotations resp. Lorentz transformations 
\be
X^a \to \L^a_b X^b, \qquad \L^a_b \in SO(D-1,1).
\label{lorentz}
\ee
These symmetries 
lead to conservation laws according to Noethers theorem,
which are elaborated below for the case of translations; see also 
\cite{Douglas:2001ba} for a related discussion.
Adapting a standard trick, we consider the following
non-constant infinitesimal transformation
$X^a \to X^a + \d X^a$ for
\be
\d X^a = \{X^b,[X^a,\varepsilon^{b'}]\}g_{bb'}
\label{transl-diffeo}
\ee
where $\varepsilon^b$ is an arbitrary matrix, and
$g_{ab} = \d_{ab}$ or $g_{ab} = \eta_{ab}$.
As elaborated in Appendix C, this leads to
\be
\frac 14 \d S_{\rm YM} 
= - Tr  \varepsilon^c [X^a,\tilde T^{a'c'}] g_{aa'} g_{cc'}
\label{T-cons}
\ee
for arbitrary $\varepsilon^a$, where 
\be
\tilde T^{ab} =  [X^a,X^c][X^b,X^{c'}]g_{cc'} 
  + [X^b,X^c][X^a,X^{c'}] g_{cc'}
  - \frac 12 g^{ab} [X^d,X^c][X^{d'},X^{c'}] g_{dd'} g_{cc'}
\label{e-m-tens}
\ee
is the  matrix - ``energy-momentum tensor''.
Since \eq{T-cons} vanishes on-shell, the conservation law 
\be
[X^a,\tilde T^{a'c}] g_{aa'} =0
\label{e-m-cons}
\ee
follows. This can of course 
also be checked directly using $[X^a,[X^b,X^{a'}]] g_{aa'} =0$. 
Moreover, since it is a consequence of a 
symmetry of the action, this will survive quantization in the form of 
a Ward identity.
Indeed it is easy to check that \eq{transl-diffeo} 
defines  a measure-preserving vector field
on the space of matrices $X^a$, so that \eq{e-m-cons}
also holds under the matrix path integral i.e. upon quantization;
there will be additional terms in the presence of matter
or in correlators.
Note that the indices of the ``tensor'' $\tilde T^{ab}$ 
range from $1$ to $D$, including transversal components.
A covariant form of these conservation laws 
and their physical meaning in the context of gravity
remains to be elaborated. 

A very similar conserved energy-momentum tensor was obtained
previously in \cite{AbouZeid:2001up,Das:2002jd} in the context of 
NC gauge theory on the Moyal-Weyl quantum plane. In that case, it was
possible to find a suitable gauge invariant version of $\tilde T^{ab}$
which satisfies a standard conservation law \cite{AbouZeid:2001up}.
The present result is somewhat different since \eq{e-m-cons} is obtained
for NC spaces with general $\theta^{\mu\nu}(x)$, involving
also components which are transversal to the brane. Moreover, the
meaning of gauge invariance versus locality is somewhat different (and
not entirely clear) in the present context; for example, $U(1)$ gauge
transformations are now interpreted as symplectomorphisms.
In any case, a similar ``local'' version of \eq{e-m-tens} involving 
Wilson lines might help to clarify its interpretation.
An analogous energy-momentum tensor in the context
of the BFSS matrix model was also found 
in \cite{Okawa:2001if}.

\section{Discussion and outlook}

We present in this paper a general framework for studying emergent
gravity in the context of Yang-Mills type matrix models,
on generic noncommutative branes embedded in $\R^D$.
The basic message is that the dynamics of fields 
on the brane is governed by an effective metric 
in the semi-classical limit, which depends
both on the embedding and the Poisson or noncommutative structure
on the brane. 
The resulting geometry is dynamical, governed by the 
matrix model and its induced effective action which includes in
particular the Einstein-Hilbert term. Therefore Yang-Mills matrix
models contain some type of gravity theory.
The results of \cite{Steinacker:2007dq} are thus generalized
to a richer class of geometries, setting
the stage for a systematic exploration of the physical properties of the 
models. This necessity to consider nontrivially embedded branes in
higher dimensions
is shown by a detailed analysis of the Newtonian limit of the 
$D=4$ model, which does not
correctly reproduce the relativistic corrections 
to the Newtonian limit.

Matrix models such as the IKKT model therefore
provide a simple and transparent
mechanism for gravity, which arises from fluctuations of the basic
matrix degrees of freedom, along with nonabelian gauge fields.
While the IKKT model was proposed originally as
non-perturbative description of IIB string theory 
\cite{Ishibashi:1996xs,Aoki:1998vn}, 
the progress in this and related works shifts the emphasis 
towards the consideration of general {\em noncommutative} 
branes and geometries, which promise to provide the physically 
relevant backgrounds. They appear to be simpler and more natural
in this context than classical spaces and geometries, the 
essential difference being the effective metric
which involves the noncommutative resp. Poisson structure.
Similar considerations should apply also to 
time-dependent matrix models such as the BFSS model 
\cite{Banks:1996vh}.

There are some important differences to General Relativity. 
The essential point is that the metric is not a
fundamental degree of freedom, but  arises effectively as 
described above.
This leads to important simplifications for the 
quantization: first, the issue of gauge fixing is much
simpler, involving degrees of freedom which can be viewed as 
scalar and gauge fields in a NC background. 
Second, it is not the Einstein-Hilbert action which is quantized,
rather the matrix model action, which is similar to a 
Yang-Mills action.
This allows to compute e.g. the one-loop effective action in a
straightforward
way, which boils down to computations in a NC gauge theory or the 
use of standard heat-kernel expansions under certain conditions.
Most remarkably, in the case of maximal supersymmetry (i.e. the
IKKT model in $D=10$) the model can be expected to be finite, 
leading to the identification of  the Planck scale 
with the scale of $N=4$ SUSY breaking. 
This suggests that the IKKT model may provide a well-defined quantum theory
of fundamental interactions including gravity.

Remarkably, emergent NC gravity appears to
provide a mechanism for avoiding
the cosmological constant problem, which is explained in the 
case of 4-dimensional branes. Again, the full significance
can only be judged once near-realistic solutions are found 
and understood.
Here the compactification of higher-dimensional
NC branes as indicated may turn out to be important, 
which is motivated also from particle physics, providing a mechanism for 
gauge symmetry breaking and fermionic zero modes.
A full discussion of emergent gravity in such cases 
is a challenging subject for future work.

This paper contains only semi-classical considerations. These 
are the leading terms in a systematic expansion 
in $\theta^{\mu\nu}$, which should be elaborated eventually. 
This can be achieved using the Seiberg-Witten map \cite{Seiberg:1999vs},
which allows to systematically re-write a noncommutative (gauge) theory 
in terms of a commutative one. While this was used in 
\cite{Steinacker:2007dq}  to 
obtain the semi-classical limit of the nonabelian gauge fields 
in emergent gravity, it is not part of the definition of the
model: it is simply -- by definition -- a natural
way to extract the physical content of a NC model. 
In principle, the quantization 
should be done on the level of the matrix model, and its
effective action can then be interpreted in a commutative language. 
For example, the issue of UV/IR mixing is resolved here not through the
Seiberg-Witten map but through its proper interpretation 
in terms of gravity \cite{Steinacker:2007dq,Klammer:2008df}.
At least in the case of (softly broken) $N=4$ SUSY, one may hope 
to resolve similarly the issues of unitarity and Wick rotation. 
All this clearly requires much more work.

Let us summarize the main arguments supporting 
emergent NC gravity as described by $D$-dimensional matrix models:
\bit
\item
The models do describe some gravity theory on 4-dimensional NC branes,
since matter couples to a universal metric 
(up to possibly density factors). Gauge fields and gravity are 
naturally unified.
\item
The class of geometries is rather rich in the case of 
models with $D>4$.
\item
The geometry is dynamical, governed by an effective action which includes the 
Einstein-Hilbert term at the quantum level. The quantization is 
likely to be well-defined, at least for the IKKT model.
\item
Flat space is a solution even at the quantum level, without fine-tuning
\item
The models are extremely simple, without any classical-geometric 
prerequisites.
\eit
This certainly describes a very promising theory
of gravity,
the main missing item being the analog of the Schwarzschild solution.
This requires to consider nontrivial embedding as shown here, 
and is complicated by the fact that the quantum effective action is 
required at least at one loop.

\paragraph{Acknowledgments}

I would like to thank 
M. Buric, C-S. Chu, L. Freidel, H. Grosse, J. Madore, 
I. Sachs, P. Schupp,  L. Smolin for useful discussions. 
This work was supported by the FWF project P20017.

\section*{Appendix A: Some identities}

The following is an
important identity for Poisson tensors:
\bea
\partial_\mu \theta^{\mu\nu} &=& 
-\theta^{\mu\mu'}\partial_\mu \theta^{-1}_{\mu'\nu'} \theta^{\nu'\nu}  \nn\\
&=& \theta^{\mu\mu'}\theta^{\nu'\nu} 
(\partial_{\mu'} \theta^{-1}_{\nu'\mu} 
 + \partial_{\nu'} \theta^{-1}_{\mu\mu'})  \nn\\
&=& - \theta^{\mu\mu'}\partial_{\mu'}\theta^{\nu'\nu}  \theta^{-1}_{\nu'\mu}
- \theta^{\nu'\nu} \partial_{\nu'}\theta^{\mu\mu'}\theta^{-1}_{\mu\mu'}  \nn\\
&=& - \partial_{\mu'}\theta^{\mu'\nu}  
- 2\theta^{\nu'\nu} \rho^{-1}\partial_{\nu'}\rho
\eea
noting that  $2\rho^{-1}\partial_{\nu}\rho = \partial_{\nu}\theta^{\mu\mu'}\theta^{-1}_{\mu\mu'}$,
hence
\be \fbox{$
\partial_\mu (\rho\,\theta^{\mu\nu}) \equiv 0  \,. $}
\label{partial-theta-id}
\ee

\paragraph{ On-shell vanishing of $\tilde \Gamma^\mu$:}

For our restricted class of metrics,
the above identity 
\eq{partial-theta-id} together with 
$|\tilde G_{\mu\nu}|^{1/2} = \rho e^{\sigma}$  
\eq{rho-a-id} implies
\bea
\tilde\Gamma^\mu &=& - |\tilde G_{\rho\sigma}|^{-1/2}
\partial_\nu (\tilde G^{\nu\mu}\,|\tilde G_{\rho\sigma}|^{1/2}) 
= - \frac 1\rho e^{-\sigma}\,
\partial_\nu (G^{\nu\mu}\,\rho) \nn\\
&=&  - \frac 1\rho e^{-\sigma}\,
\partial_\nu (\rho\,\theta^{\nu\nu'}\theta^{\mu\mu'}
 g_{\mu'\nu'}(x)) \nn\\ 
&=&  - e^{- \sigma}\,\theta^{\nu\nu'}
\partial_\nu (\theta^{\mu\mu'} g_{\mu'\nu'}(x)) 
\,\,\stackrel{\rm e.o.m.}{=}\,\, 0
\label{tilde-Gamma-vanish}
\eea
using the e.o.m. \eq{eom-extradim} for $X^\mu$.
This can also be seen from \eq{eom-varphi}.
Therefore the equations of motion for $X^\mu$ are 
equivalent to $\tilde\Gamma^\mu=0$. From the 
point of view of General Relativity, this would be interpreted
rather as a gauge-fixing condition.
This is not the case here due to the constrained class of metrics.

\section*{Appendix B:
Derivation of the covariant e.o.m.}

Consider
\bea
\tilde G^{\gamma\eta}(x)\, \tilde\nabla_\gamma \theta^{-1}_{\eta\nu} &=&
\tilde G^{\gamma\eta}(x)\,\(\partial_\gamma \theta^{-1}_{\eta\nu} 
- \tilde \Gamma_{\gamma\eta}^\rho \theta^{-1}_{\rho\nu}
- \tilde \Gamma_{\gamma\nu}^\rho \theta^{-1}_{\eta \rho}\) \nn\\
&=& \tilde G^{\gamma\eta }\,\partial_\gamma \theta^{-1}_{\eta \nu} 
- \tilde \Gamma^\rho \theta^{-1}_{\rho\nu}
- \tilde G^{\gamma\eta }\,\tilde \Gamma_{\gamma\nu}^\rho \theta^{-1}_{\eta \rho} 
\eea
where 
\be
\tilde \Gamma^\gamma = \tilde G^{ab}\tilde \Gamma^\gamma_{ab} = 
- \frac 1{\sqrt{\tilde G_{ab}}} 
\partial_\rho(\tilde G^{\gamma \rho} \sqrt{\tilde G_{ab}}) \, .
\label{Gamma-def}
\ee
Using \eq{G-tilde-general}
\be
\tilde G^{\mu\nu} = e^{-\sigma}\, G^{\mu\nu}
 =  \theta^{\mu\mu'}(x) \theta^{\nu\nu'}(x) \, \tilde g_{\mu'\nu'}(x) 
\ee
where
\be
\tilde g_{\mu'\nu'}(x) \equiv  e^{-\sigma}\,g_{\mu'\nu'}(x), 
\label{g-tilde-closed}
\ee
we can write 
\bea
\tilde G^{\gamma\eta}\,\tilde \Gamma_{\gamma\nu}^\delta \theta^{-1}_{\eta \delta}
&=& \frac 12 \tilde G^{\gamma\eta}\, \tilde G^{\rho \delta} \theta^{-1}_{\eta \delta}
\(\partial_\gamma \tilde G_{\rho\nu}+ \partial_\nu \tilde G_{\rho\gamma} - \partial_\rho \tilde G_{\gamma\nu} \)\nn\\
 &=& \frac 12  \tilde\theta^{\gamma \rho}
 \(\partial_\gamma \tilde G_{\rho\nu}+ \partial_\nu \tilde G_{\rho\gamma} - \partial_\rho \tilde G_{\gamma\nu} \)\nn\\
&=& \tilde\theta^{\gamma\rho}\partial_\gamma \tilde G_{\rho\nu}
 =  \tilde G^{\gamma\eta} \tilde G^{\rho \delta}\, \theta^{-1}_{\eta \delta}\,\partial_\gamma \tilde G_{\rho \nu}  \nn\\
&=& - \tilde G^{\gamma\eta} (\partial_\gamma \tilde G^{\rho \delta})\, \theta^{-1}_{\eta \delta}\, \tilde G_{\rho \nu}  \nn\\
&=& - \tilde G^{\gamma\eta} \theta^{-1}_{\eta \delta}\, \tilde G_{\rho \nu} 
(\partial_\gamma (\theta^{\rho \mu}\tilde g_{\mu \delta'}) \theta^{\delta\delta'} 
 + \theta^{\rho \mu} \tilde g_{\mu \delta'}\partial_\gamma\theta^{\delta\delta'}) \nn\\
&=& - \tilde G^{\gamma\eta} \tilde G_{\rho \nu}\, \tilde g_{\mu \eta}\partial_\gamma \theta^{\rho \mu }
  - \tilde G^{\gamma\eta} \theta^{\rho \mu } \tilde G_{\rho \nu} \partial_\gamma \tilde g_{\mu \eta}
+ \tilde G^{\gamma\eta}\partial_\gamma\theta^{-1}_{\eta\nu} 
\eea
since $\tilde\theta^{\gamma \rho } := \tilde G^{c\eta} \tilde G^{\rho \delta}\, \theta^{-1}_{\eta \delta}$
is antisymmetric. 
The Jacobi identity gives
\bea
\tilde G^{\gamma\eta} \tilde g_{\eta \mu }\, \partial_\gamma 
\theta^{\rho\mu } &=&
\theta^{\eta \eta'} \tilde g_{\gamma'\eta'} \,\tilde g_{\eta \mu }  \theta^{\gamma\gamma'} \partial_\gamma \theta^{\rho \mu } \nn\\
&=&\theta^{\eta\eta'} \tilde g_{\gamma'\eta'} \, \tilde g_{\eta \mu } 
(\theta^{\rho\gamma} \partial_\gamma \theta^{\mu \gamma'} + \theta^{\mu \gamma} \partial_\gamma \theta^{\gamma'\rho}) \nn\\
&=& \theta^{\rho\gamma} \tilde g_{\gamma'\eta'}  \theta^{\eta\eta'}\tilde g_{\eta \mu }\,  
\partial_\gamma \theta^{\mu \gamma'}
+ (\theta^{\eta\eta'} \theta^{\mu \gamma} \tilde g_{\eta \mu }) \tilde g_{\gamma'\eta'} \,  \partial_\gamma \theta^{\gamma'\rho} \nn\\
&=& \theta^{\rho\gamma} \tilde g_{\gamma'\eta'} \theta^{\eta\eta'}\tilde g_{\eta \mu }
 \partial_\gamma \theta^{\mu \gamma'}
- \tilde G^{\gamma\eta} \tilde g_{\mu \eta} \, \partial_\gamma \theta^{\rho\mu } 
\eea
hence
\bea
\tilde G^{\gamma\eta} \tilde g_{\eta \mu }\, 
\partial_\gamma \theta^{\rho\mu } 
&=& \frac 12 \theta^{\rho\gamma} \tilde g_{\gamma'\eta'} \theta^{\eta\eta'}\tilde g_{\eta \mu } \partial_\gamma \theta^{\mu \gamma'} \,.
\eea
Finally, observe that
\bea
G^{\gamma\eta }\,\partial_\gamma \d g_{\mu\eta } - G^{\gamma\eta }\,\partial_\mu \d g_{\gamma\eta }
&=& G^{\gamma\eta }\,\partial_\gamma (\partial_\mu\phi\partial_\eta \phi) 
    - G^{\gamma\eta }\,\partial_\mu (\partial_\gamma\phi\partial_\eta \phi) \nn\\
&=& \partial_\mu\phi \, G^{\gamma\eta }\,\partial_\gamma \partial_\eta \phi 
 + G^{\gamma\eta }\,\partial_\mu\partial_\gamma\phi \partial_\eta \phi 
 - G^{\gamma\eta }\,\partial_\mu \partial_\gamma\phi\partial_\eta \phi
 - G^{\gamma\eta }\,\partial_\gamma\phi\partial_\mu \partial_\eta \phi \nn\\
&=& \partial_\mu\phi\,  G^{\gamma\eta }\,\partial_\gamma \partial_\eta \phi 
 - \frac 12 G^{\gamma\eta }\,\partial_\mu \d g_{\gamma\eta }
\eea
hence
\be
G^{\gamma\eta }\,\partial_\gamma \d g_{\mu\eta } 
= \frac 12 G^{\gamma\eta }\,\partial_\mu \d g_{\gamma\eta }
 + \partial_\mu\phi\,  G^{\gamma\eta }\,\partial_\gamma \partial_\eta \phi 
\ee
and
\be
\tilde G^{\gamma\eta }\,\partial_\gamma  g_{\mu\eta } 
= \frac 12 \tilde G^{\gamma\eta }\,\partial_\mu  g_{\gamma\eta }
 + \partial_\mu\phi\,  \tilde G^{\gamma\eta }\,\partial_\gamma \partial_\eta \phi \,.
\ee
Therefore
\be
\tilde G^{\gamma\eta }\,\partial_\gamma \tilde g_{\mu\eta } 
= \frac 12 \tilde G^{\gamma\eta }\,\partial_\mu \tilde g_{\gamma\eta }
 + e^{-\sigma}\,\partial_\mu\phi\, \tilde G^{\gamma\eta }\,\partial_\gamma \partial_\eta \phi 
- \tilde G^{\gamma\eta } \tilde g_{\mu\eta }\,\partial_\gamma \sigma
 + \frac 12 \tilde G^{\gamma\eta } \tilde g_{\gamma\eta }\, \partial_\mu  \sigma
\ee
and we obtain
\bea
\tilde G^{\gamma\eta } \tilde g_{\eta \mu}\, \partial_\gamma \theta^{\rho\mu} +
\tilde G^{\gamma\eta } \theta^{\rho\mu} \partial_\gamma \tilde g_{\mu\eta }
&=& \frac 12 \theta^{\rho\gamma} \tilde g_{\gamma'\eta '} \theta^{\eta \eta '}\tilde g_{\eta \mu} \partial_\gamma \theta^{\mu\gamma'}
+\frac 12 \theta^{\rho\gamma}\,\tilde G^{\mu\eta }\partial_\gamma \tilde g_{\mu\eta } \nn\\
&& + \theta^{\rho\mu}\,(e^{-\sigma}\,\partial_\mu\phi\, \tilde G^{\gamma\eta }\partial_\gamma \partial_\eta \phi 
 - \tilde G^{\gamma\eta } \tilde g_{\mu\eta }\,\partial_\gamma \sigma
 + \frac 12 \tilde G^{\gamma\eta } \tilde g_{\gamma\eta }\, \partial_\mu \sigma) \nn\\
&=& \theta^{\rho\mu}\,(\partial_\mu\tilde\eta(x) + e^{-\sigma}\,\partial_\mu\phi\, \tilde G^{\gamma\eta }\partial_\gamma \partial_\eta \phi 
 - \tilde G^{\gamma\eta } \tilde g_{\mu\eta }\,\partial_\gamma \sigma
 + 2 \tilde\eta(x)\, \partial_\mu \sigma)\nn
\eea
using the scalar function 
\be
\tilde\eta(x) = \frac 14 \tilde G^{\mu\nu} \tilde g_{\mu\nu} 
=  \frac 14 \theta^{\mu\mu'} \tilde g_{\mu'\nu'} \theta^{\nu\nu'}\tilde g_{\mu\nu} 
\ee
which satisfies
\be
\partial_\gamma \tilde\eta(x) =  \frac 12 \partial_\gamma\theta^{\mu\mu'} \tilde g_{\mu'\nu'} \theta^{\nu\nu'}\tilde g_{\mu\nu}  
 +  \frac 12 \tilde G^{\mu\nu} \partial_\gamma \tilde g_{\mu\nu} \,.
\ee
Putting all this together, we obtain
\bea
\tilde G^{\gamma\eta }(x)\, \tilde\nabla_\gamma \theta^{-1}_{\eta \nu}
&=& \tilde G^{\gamma\eta }\,\partial_\gamma \theta^{-1}_{\eta \nu} 
- \tilde G^{\gamma\eta }\,\tilde \Gamma_{\gamma\nu}^\mu \theta^{-1}_{\eta \mu}
- \tilde \Gamma^\mu \theta^{-1}_{\mu\nu} \nn\\
&=& \tilde G_{\rho\nu}\,\(\tilde G^{\gamma\eta } \tilde g_{\mu\eta }\partial_\gamma \theta^{\rho \mu}
  + \tilde G^{\gamma\eta } \theta^{\rho \mu} \partial_\gamma \tilde g_{\mu\eta } \)
 - \tilde \Gamma^\mu \theta^{-1}_{\mu\nu} \nn\\
&=& \tilde G_{\rho\nu}\,\(\theta^{\rho \mu} (\partial_\mu\tilde\eta(x)
  + e^{-\sigma}\,\partial_\mu\phi\, 
\tilde G^{\gamma\eta }\partial_\gamma \partial_\eta \phi 
 - \tilde G^{\gamma\eta } \tilde g_{\mu\eta }\,\partial_\gamma \sigma
 + 2 \tilde\eta(x)\, \partial_\mu \sigma) \)
- \tilde \Gamma^\mu \theta^{-1}_{\mu\nu}\,. \nn
\eea
This holds identically, i.e. it 
characterizes the {\em constraint} of the metric.

Now we take into account the equations of motion
$\tilde \Gamma^\mu =0$ \eq{tilde-Gamma-vanish} and 
$\Delta_{\tilde G} \phi 
= \tilde G^{\mu\nu} \partial_\mu\partial_\nu\phi =0$ 
\eq{eom-phi}, 
which hold in the special coordinates $x^\mu$ defined by the 
dynamical matrices (this is why the above is non-covariant). 
We can then rewrite this as
\bea
\tilde G^{\gamma\eta}(x)\, \tilde\nabla_\gamma \theta^{-1}_{\eta\nu}
&=& \tilde G_{\rho\nu}\theta^{\rho\mu}\,\( \partial_\mu\tilde\eta(x)
 - \tilde G^{\gamma\eta} \tilde g_{\mu\eta}\,\partial_\gamma \sigma
 + 2 \tilde\eta(x)\, \partial_\mu \tilde \sigma \) \nn\\
&=& \tilde G_{\rho\nu}\theta^{\rho\mu}\,\( \partial_\mu\tilde\eta(x)
 + 2 \tilde\eta(x)\, \partial_\mu \sigma \)
 - \tilde G^{\gamma\eta}\theta^{-1}_{\eta\nu} \,\partial_\gamma \sigma \,.
\label{theta-covar}
\eea
Using 
$\eta = e^{2\sigma}\tilde\eta(x)
 = \frac 14 \theta^{\mu\mu'} g_{\mu'\nu'} \theta^{\nu\nu'} g_{\mu\nu}$ \eq{eta-def},
we obtain
\be
\tilde G^{\gamma\eta}(x)\, \tilde\nabla_\gamma 
(e^{\sigma}\, \theta^{-1}_{\eta\nu})
= e^{\sigma}\tilde G_{\rho\nu}\theta^{\rho \gamma}\,
\( \partial_\gamma\tilde \eta(x)
 + 2 \tilde \eta(x)\, \partial_\gamma \sigma \) 
= e^{-\sigma}\,\tilde G_{\rho\nu}\theta^{\rho \gamma}\,
\partial_\gamma\eta(x) 
\ee
which is \eq{eom-geom-covar-1} resp. \eq{eom-geom-covar-extra}.
This is the covariant form of the equation of motion, 
independent of the choice of coordinates.

\section*{Appendix C:
Derivation of \eq{e-m-cons}}

We use here a short-hand notation where double upper
indices are understood to be contracted with $\d_{ab}$
or $\eta_{ab}$. Then
\bea
0 &=& -\frac 14 \d S_{\rm YM} = Tr[\d X^a,X^b][X^a,X^b] \nn\\
 &=& Tr[\{X^c,[X^a,\varepsilon^c]\},X^b][X^a,X^b] \nn\\
&=& Tr[X^c [X^a,\varepsilon^c] + [X^a,\varepsilon^c] X^c,X^b][X^a,X^b] 
 \nn\\
 &=& Tr \Big( X^c [[X^a,\varepsilon^c],X^b][X^a,X^b]
  + [[X^a,\varepsilon^c],X^b] X^c [X^a,X^b] \nn\\
 && + [X^a,\varepsilon^c] [X^c,X^b][X^a,X^b]
 +  [X^c,X^b] [X^a,\varepsilon^c] [X^a,X^b]\Big) \nn\\
&=& Tr \Big( \{X^c, [[X^a,\varepsilon^c],X^b]\} [X^a,X^b]
 + [X^a,\varepsilon^c] \{[X^c,X^b],[X^a,X^b]\}\Big) \nn\\
&=&  Tr \Big( -\{X^c, [X^b,[X^a,\varepsilon^c]]\} [X^a,X^b]
 + [X^a,\varepsilon^c] \{[X^c,X^b],[X^a,X^b]\}\Big) \nn\\
&=& Tr \Big( \frac 12\{X^c, [\varepsilon^c,[X^b,X^a]]\} [X^a,X^b]
 + [X^a,\varepsilon^c] \{[X^c,X^b],[X^a,X^b]\}\Big)
\eea
for arbitrary $\varepsilon^a$.
Using $Tr(\{A,[B,C]\}C) = Tr([A,B] C^2)$ this can be written as
\bea
0 &=& Tr \Big(- \frac 12[X^c,\varepsilon^c][X^a,X^b][X^a,X^b]
 + [X^a,\varepsilon^c] \{[X^c,X^b],[X^a,X^b]\}\Big) \nn\\
&=& Tr \frac 12 \varepsilon^c [X^c,[X^a,X^b][X^a,X^b]]
  - \varepsilon^c [X^a, \{[X^c,X^b],[X^a,X^b]\}]\Big) \nn\\
 &=& Tr  \varepsilon^c [X^a,\tilde T^{ac}] \,.
\eea

\end{document}